\documentclass[aps,prd,twocolumn,nobalancelastpage,superscriptaddress,showpacs]{revtex4}

\usepackage{graphicx}
\usepackage{amsmath}
\usepackage{float}
\usepackage{nicefrac}
\usepackage{multirow}
\usepackage{color}
\usepackage[colorlinks]{hyperref}

\newcommand{\beq}{\begin{equation}}
\newcommand{\eeq}{\end{equation}}

\newcommand{\gianfranco}[1]{}
\newcommand{\pierre}[1]{}
\newcommand{\pierreb}[1]{}
\newcommand{\citeeq}[1]{Eq.~(\ref{#1})}
\newcommand{\citefig}[1]{Fig.~\ref{#1}}
\begin{document}
%
\title{Antiproton and Positron Signal Enhancement in Dark Matter Mini-Spikes Scenarios}
\author{Pierre Brun}
\email{brun@lapp.in2p3.fr}
\affiliation{Laboratoire d'Annecy-le-Vieux de
Physique des Particules LAPP, Universit\'e de Savoie, CNRS/IN2P3, 74941
Annecy-le-vieux, France}
\author{Gianfranco Bertone}
\email{bertone@iap.fr} \affiliation{INFN, Sezione di Padova, Via Marzolo 8,
Padova I-35131, Italy} \affiliation{Institut d'Astrophysique de Paris, UMR
7095-CNRS, Universit\'e Pierre et Marie Curie, 75014 Paris, France}
\author{Julien Lavalle}
\email{lavalle@in2p3.fr}
\affiliation{Centre de Physique des Particules de
Marseille CPPM, Universit\'e de la M\'edit\'erann\'ee, CNRS/IN2P3, 13288
Marseille, France}
\author{Pierre Salati}
\email{salati@lapp.in2p3.fr}
\affiliation{Laboratoire d'Annecy-le-Vieux de
Physique Th\'eorique LAPTH, Universit\'e de Savoie, CNRS/IN2P3, 74941
Annecy-le-vieux, France}
\author{Richard Taillet}
\email{taillet@lapp.in2p3.fr}
\affiliation{Laboratoire d'Annecy-le-Vieux de
Physique Th\'eorique LAPTH, Universit\'e de Savoie, CNRS/IN2P3, 74941
Annecy-le-vieux, France}
\date{\today}

\begin{abstract}
The annihilation of dark matter (DM) in the Galaxy could produce specific
imprints on the spectra of antimatter species in Galactic cosmic rays, which
could be detected by upcoming experiments such as PAMELA and AMS02. Recent
studies show that the presence of substructures  can enhance the annihilation
signal by a "boost factor" that not only depends on energy, but that is
intrinsically a statistical property of the distribution of DM substructures
inside the Milky Way. We investigate a scenario in which substructures consist
of $\sim 100$ "mini-spikes" around intermediate-mass black holes. Focusing on
primary positrons and antiprotons, we find large boost factors, up to a few
thousand, that exhibit a large variance at high energy in the case of positrons
and at low energy in the case of antiprotons. As a consequence, an estimate of
the DM particle mass based on the observed cut-off in the positron spectrum
could lead to a substantial underestimate of its actual value.
\end{abstract}

\pacs{..,..,..,..,.. \hspace{2.5cm} LAPP-EXP-2007-02, LAPTH-1181/07,
CPPM-P-2007-01}

\maketitle
%
%
Many current and upcoming experiments are aimed at detecting the products of
annihilation of dark matter particles in the Milky Way halo, or in external
galaxies -- see Refs.~\cite{Bergstrom:2000pn,Munoz:2003gx,Bertone:2004pz} for
recent reviews.
In fact, the identification of an annihilation signal would not only provide
the proof of the particle nature of DM, but it would also open a new window to
Physics beyond the Standard Model. Among indirect searches, particular
attention has been devoted to the case of photons, since they propagate along
straight lines, and they are not significantly absorbed within the galaxy at
energies of order 100 GeV and below. The prospects for detecting gamma-rays
from DM annihilation with the space telescope GLAST~\cite{glast} and with Air
Cherenkov telescopes such as CANGAROO~\cite{canga}, HESS~\cite{hess},
MAGIC~\cite{magic} and VERITAS~\cite{veritas}, have been extensively discussed
in literature -- e.g.~\cite{Bertone:2005xv} and references therein.
Alternatively, one may search for neutrino fluxes from DM captured at the
center of the Sun or of the Earth -- see reviews above for further details --
with large neutrino telescopes such as IceCube~\cite{icecube} and
Antares~\cite{Aslanides:1999vq}.

One can also focus on other final states of DM annihilation, such as antimatter
particles, in particular positrons
\cite{positrons1,positrons2,Baltz:2001ir,positrons3,positrons4,
positrons5,positrons6,Cheng:2002ej,Lavalle:2006} and
antiprotons~\cite{antiproton1,Donato:2003xg,Bergstrom:1999jc,Lionetto}. Unlike
photons and neutrinos, these are charged particles that are sensitive to the
galactic magnetic field and suffer energy losses. Their propagation must be
described by diffusion, so that the magnetic properties of the Milky Way must
be known, which is not completely true yet. Nevertheless, it is possible to
make interesting predictions, which are now particularly relevant, as the
recently launched satellite PAMELA~\cite{Picozza:2006nm} and the forthcoming
AMS02~\cite{ams02} experiment are expected to provide new data and test a
significant portion of the parameter space of DM particles.

In most cases, the prospects for the detection of antimatter fluxes are not
particularly sensitive to the overall shape of the DM profile, but they do
depend strongly on the ``clumpiness'' of the profile. In fact, the annihilation
rate being proportional to the square of the number density, the presence of
small-scale inhomogeneities leads to larger fluxes for the same value of the
average DM density. This is often parameterized by a ``boost factor'', that
will be defined more precisely below.

Recenty, a new scenario has been discussed in the literature, where the
formation of Intermediate Mass Black Holes, i.e. Black Holes with mass $M$ in
the range $10^{2} \lesssim M/{\rm M_{\odot}} \lesssim 10^{6}$, leads to the
formation of DM overdensities, called ``mini-spikes'', that might be observed
as point sources of gamma-rays~\cite{Bertone:2005xz} and
neutrinos~\cite{Bertone:2006nq}. Although somewhat speculative, the scenario
has the undisputed virtue of making specific predictions on the number and
luminosity of these objects, that could be observed or ruled out in the near
future with the upcoming generation of space and ground based experiments.
Here, we compute the boost factor in mini-spikes scenarios, and argue that they
would lead to a dramatic enhancement of antimatter fluxes, bringing them within
the reach of current and upcoming experiments.

The paper is organized as follows~: in Section~\ref{sec:the_scenario}, we
review the motivations for the mini-spike scenario and derive their density
profile and normalization.
Section~\ref{sec:the_method} is devoted to a general discussion of the
antimatter cosmic ray signal generated by mini-spikes. The average value and
the variance of the boost factor at the Earth are defined and special attention
is paid to the inner structure and galactic distribution of the mini-spikes.
Positrons and antiprotons are respectively addressed in
Sections~\ref{sec:positrons} and \ref{sec:antiprotons}. Results from
Monte-Carlo simulations are presented and discussed in the light of analytic
calculations. The energy dependence of the boost variance turns out to be quite
different for positrons and antiprotons. These species have very different
modes of propagation inside the Milky Way.
A few realistic models for the DM particles are implemented in
Section~\ref{sec:real_spectra}. The positron and antiproton signals which they
yield are featured and illustrate the typical antimatter signatures to be
expected should mini-spikes populate the galactic halo.
We finally conclude in Section~\ref{sec:conclusion} and make a few suggestions
for future developments.

\section{Mini-spikes as DM substructures}
\label{sec:the_scenario}

Black Holes (BHs) can be broadly divided in three different classes -- see
e.g.~\cite{Miller:2003sc} for a review~:
\begin{itemize}
\item {\it Stellar Mass BHs}, with mass $M \lesssim 100 \;{\rm M_{\odot}}$.
They are typically remnants of the collapse of massive stars. Recent
simulations suggest that the upper limit on the mass of these objects is as low
as $\approx 20 \;{\rm M_{\odot}}$~\cite{Fryer:2001}. There is {\it robust}
evidence for the existence of these objects, coming from the observation of
binary systems with compact members whose mass exceeds the critical mass of
Neutron Stars. For a review of the topic and the discussion of the status of
the observational evidence for Stellar Mass BHs see e.g.~\cite{Narayan:2003fy}
and references therein.
\item {\it Supermassive BHs (SMBHs)}, with mass $M \gtrsim 10^{6} \;{\rm
M_{\odot}}$ lying at the centers of galaxies, including our own. Their
existence is also well-established -- see e.g. Ref.~\cite{Ferrarese:2005} --
and intriguing correlations are observed between the SMBHs mass and the
properties of their host galaxies and
halos~\cite{kormendy:1995,Ferrarese:2000se,McLure:2001uf,Gebhardt:2000fk,
Tremaine:2002js,Koushiappas:2003zn}. From a theoretical point of view, a
population of massive seed black holes could help to explain the origin of
SMBHs. In fact, observations of quasars at redshift $z \approx 6$ in the Sloan
Digital survey~\cite{Fan:2001ff,barth:2003,Willott:2003xf} suggest that SMBHs
were already in place when the Universe was only $\sim$1 Gyr old, a
circumstance that can be understood in terms of a rapid growth starting from
``massive'' seeds -- see e.g. Ref.~\cite{haiman:2001}. This leads to the third
category:
\item {\it Intermediate Mass BHs (IMBHs)}, with mass $10^{2} \lesssim M/{\rm
M_{\odot}} \lesssim 10^{6}$. Scenarios that seek to explain the properties of
the observed supermassive black holes population result in the prediction of a
large population of wandering Intermediate Mass BHs (IMBHs). Here, following
Ref.~\cite{Bertone:2005xz}, we consider two different formation scenarios for
IMBHs.
In the first scheme (A), IMBHs form in rare, overdense regions at high
redshift, $z \sim 20$, as remnants of Population III stars, and have a
characteristic mass-scale of a few $10^{2} \, {\rm M_{\odot}}$
\cite{Madau:2001} -- a similar model was investigated in
Ref.~\cite{Zhao:2005zr,islamc:2004,islamb:2004}. In this scenario, these black
holes serve as the seeds for the growth of supermassive black holes found in
galactic spheriods \cite{Ferrarese:2005}.
In the second scenario (B), IMBHs form directly out of cold gas in
early-forming halos and are typified by a larger mass scale of order $10^{5} \,
{\rm M_{\odot}}$~\cite{Koushiappas:2003zn}. During the virialization and
collapse of the first halos, gas cools, collapses, and forms pressure-supported
disks at the centers of halos that are massive enough to contain a large amount
of molecular hydrogen. In halos which do not experience any major mergers over
a dynamical time, a protogalactic disk forms and can evolve uninterrupted. At
this stage, an effective viscosity due to local gravitational instabilities in
the disk leads to an inward mass transfer and outward angular momentum
transfer, until supernovae in the first generation of stars heat the disk and
terminate this process. By the time the process terminates, a baryonic mass of
order $\sim10^5M_{\odot}$ loses its angular momentum and is transferred to the
center of the halo, leading to the formation of an object that may be briefly
pressure-supported, but that eventually collapses to form a black hole. In
\citefig{fig:radial_dist} we show the distribution of IMBHs in the latter
scenario, as obtained in Ref.~\cite{Bertone:2006nq}.
\end{itemize}

The effect of the formation of a central object on the surrounding distribution
of matter has been investigated in
Refs.~\cite{peebles:1972,young:1980,Ipser:1987ru,Quinlan:1995} and for the
first time in the framework of DM annihilations in Ref.~\cite{Gondolo:1999ef}.
It was shown that  the {\it adiabatic} growth of a massive object at the center
of a power-law distribution of DM, with index $\gamma$, induces a
redistribution of matter into a new power-law (dubbed ``spike'') with index
\begin{equation}
\gamma_\text{sp} = (9-2\gamma)/(4-\gamma) \;\; .
\end{equation}
This formula is valid over a region of size $R_\text{sp} \approx 0.2 \,
r_{BH}$, where $r_{BH}$ is the radius of gravitational influence of the black
hole, defined implicitly as $M(<r_{BH})=M_{BH}$, where $M(<r)$ denotes the mass
of the DM distribution within a sphere of radius $r$, and where $M_{BH}$ is the
mass of the Black Hole~\cite{Merritt:2003qc}.
The process of adiabatic growth is in particular valid for the SMBH at the
galactic center.
A critical assessment of the formation {\it and survival} of the central spike,
over cosmological timescales, is presented in
Refs.~\cite{Bertone:2005hw,Bertone:2005xv} -- see also references therein.
Adiabatic spikes are rather fragile structures, that require fine-tuned
conditions to form at the center of galactic halos~\cite{Ullio:2001fb}, and
that can be easily destroyed by dynamical processes such as major
mergers~\cite{Merritt:2002vj} and gravitational scattering off
stars~\cite{Merritt:2003eu,Bertone:2005hw}.

It was recently shown that a $\rho \propto r^{-3/2}$ DM overdensity can be
predicted in any halo at the center of any galaxy old enough to have grown a
power-law density cusp {\it in the stars} via the Bahcall-Wolf
mechanism~\cite{Merritt:2006mt}. Collisional generation of these DM ``crests"
-- Collisionally REgenerated STructures -- was demonstrated even in the extreme
case where the DM density was lowered by slingshot ejection from a binary
supermassive black hole. However, the enhancement of the annihilation signal
from a DM crest is typically much smaller than for adiabatic
spikes~\cite{Merritt:2006mt}.

Here we focus our attention on {\it mini-spikes} around IMBHs, and we recall
their properties, following closely Ref.~\cite{Bertone:2005xz}. The ``initial''
DM mini-halo -- that is, the DM distribution prior to black hole formation --
can be well approximated with a Navarro, Frenk, and White (NFW)
profile~\cite{Navarro:1996he}
\begin{equation}
\rho(r) = \rho_0 \left( \frac{r}{r_\text{sc}} \right)^{-1} \left( 1 +
\frac{r}{r_\text{sc}} \right)^{-2} \;\; . \label{eq:nfw}
\end{equation}
The normalization constant $\rho_0$, and the scale radius $r_\text{sc}$, can be
expressed in terms of the virial mass $M_{{\rm vir}}$ of the halo at the time
when the IMBH formed, and of the virial concentration parameter  $c_{{\rm
vir}}$ -- see Ref.~\cite{Bertone:2005xz} for futher details.
Alternatively, we could have chosen the more recent parameterization proposed
by Navarro et al.~\cite{Navarro:04b} -- see also
Refs.~\cite{Reed:05,Merritt:05}. However, this profile implies modifications at
scales smaller than those we are interested in, where the profile is anyway
modified by the presence of the IMBH.
We assume that the black holes form over a timescale long enough to guarantee
adiabaticity, but short compared to the cosmological evolution of the host
halo.  In fact, the condition of ``adiabaticity'', fundamental to grow
mini-spikes, requires that the formation time of the black hole is much larger
than the dynamical timescale at a distance $r_{BH}$ from the black hole, where
$r_{BH} \simeq G M_{BH} / \sigma^{2}$ is the radius of the sphere of
gravitational influence of the black hole, and $\sigma$ is the velocity
dispersion of DM particles at $r_{BH}$. In practice, $r_{BH}$ is estimated by
solving the implicit equation
\begin{equation}
M(<r_{BH}) \equiv \int_{0}^{r_{BH}} \rho_{\rm tot}(r) \, 4 \pi \, r^2 \, {\rm
d}r = 2 \, M_{BH} \;\; .
\end{equation}
In the following, we only consider the formation scenario B, which leads to
higher values for the boost factor.
For a representative case in scenario B, with $M_{BH} = 10^{5} \, {\rm
M_{\odot}}$ and $M_{{\rm vir},f} = 10^{8} \, {\rm M_{\odot}}$, this gives
$r_{BH} / r_\text{sc} \approx 0.04$. In scenario B, the black hole formation
time is set by the timescale for viscous angular momentum loss and is limited
by the evolutionary timescale of the first stars and the gravitational infall
time across the gaseous disk, which is of order~Myr -- see
Ref.~\cite{Koushiappas:2005qz} for a detailed discussion of timescales. The
relevant timescale for the mass build up of the IMBH is then $t_{\rm ev} \sim 1
- 20$~Myr.

\section{The antimatter cosmic ray signal produced by mini-spikes}
\label{sec:the_method}

\subsection{Fluxes at the Earth}

The self-annihilations of the DM particles $\chi$ concealed in the Milky Way
halo produce positrons and antiprotons, a small fraction of which may reach the
Earth. The local production rate of these antimatter cosmic ray species will be
generically denoted by
\beq {\cal P}(\mathbf{x}) = \delta \, \langle \sigma_{\rm ann} v \rangle \,
\left\{ {\displaystyle \frac{\rho_{\chi}(\mathbf{x})}{m_{\chi}}} \right\}^{2}
\, g(E_{S}) \, \Delta E_{S} \;\; . \label{generic_source} \eeq
The factor $\delta$ is equal to $1/2$ if the DM particles $\chi$ are Majorana
fermions whereas it is $1/4$ in the case of Dirac species, provided that a
perfect matter-antimatter symmetry between $\chi$ and $\bar{\chi}$ holds in
that case.
Until Section~\ref{sec:real_spectra}, we will focus our discussion on
monoenergetic positrons and antiprotons with energy $E_{S}$ defined up to an
energy bandwidth of $\Delta E_{S}$. In Kaluza-Klein theories of dark matter
particles, the branching ratio into an $e^{+} e^{-}$ pair can be as large as
$B_{e^{\pm}} \sim 20$ \% \cite{appel} and the  energy distribution $g(E_{S})$
of \citeeq{generic_source} is equal to $B_{e^{\pm}} \, \delta (E_{S} -
m_{\chi})$, where $m_{\chi}$ stands for the mass of the DM species. In the case
of antiprotons, energy losses may be neglected and we need only to consider a
specific value of the initial energy $E_{S}$. A continuous injection spectrum
will be considered in Section~\ref{sec:real_spectra}, as is needed for
realistic models of antimatter production.

The probability for a cosmic ray species injected at $\mathbf{x}$ to propagate
and reach the Earth with energy $E$ is described by the Green function
\beq G(\mathbf{x}) \equiv G(E , {\odot} \leftarrow E_{S} , \mathbf{x}) \;\; .
\eeq
The propagator $G(\mathbf{x})$ has been thouroughly discussed in the literature
-- see Refs.~\cite{positrons6,Lavalle:2006} for positrons and
Refs.~\cite{G_pbar_0609522,Bringmann:2006im} for antiprotons.
The annihilations of DM particles yield the following flux at the Earth
\beq \phi(E) = {\mathcal S} \, {\displaystyle \int}_{\rm DM \, halo} \!\!
G(\mathbf{x}) \, \left\{ {\displaystyle
\frac{\rho_{\chi}(\mathbf{x})}{\rho_{0}}} \right\}^{2} \, d^{3} \mathbf{x} \;\;
, \eeq
which is expressed in units of cm$^{-2}$ s$^{-1}$ sr$^{-1}$ GeV$^{-1}$. The
factor ${\mathcal S}$ encodes informations on the specific model selected to
describe the DM species $\chi$ such as its mass $m_{\chi}$ and total
annihilation cross section $\langle \sigma_{\rm ann} v \rangle$. It is defined
as
\beq {\mathcal S} = {\displaystyle \frac{\delta}{4 \pi}} \, \beta(E) \, \langle
\sigma_{\rm ann} v \rangle \, \left\{ {\displaystyle \frac{\rho_{0}}{m_{\chi}}}
\right\}^{2} \, g(E_{S}) \, \Delta E_{S} \;\; . \label{Sfactor} \eeq
The velocity of the cosmic ray particles is denoted by $\beta(E)$ and
$\rho_{0}$ is a reference density that can be chosen at will, eventually
disappearing from the final result. It can be given a physical meaning, as we
will see when we define the annihilation volume $\xi$.

\subsection{The framework of the statistical analysis}

In the absence of any substructure, the DM distribution $\rho_{s}$ is smooth
and yields a cosmic ray flux
\beq \phi_{s}(E) = {\mathcal S} \, \int_\text{DM \, halo} \!\! G(\mathbf{x}) \,
\left\{  \frac{\rho_s(\mathbf{x})}{\rho_{0}} \right\}^{2} \, d^{3} \mathbf{x}
\;\; . \eeq
In the IMBHs scenario, the DM distribution is given by the superposition $\rho
= \rho_{s} + \delta \rho$ where a new component $\delta \rho \gg \rho_{s}$
accounts now for the DM trapped inside the mini-spikes. That component
contributes for a fraction of $\sim 10^{-5}$ to the Galactic dark matter. As
the typical distance over which positrons and antiprotons propagate in the
Milky Way is much larger than the average size of the mini-spikes, the cosmic
ray flux -- generated by the total density $\rho$ -- simplifies into the sum
\beq \phi(E) = \phi_{s} + \left( \phi_{r} = {\displaystyle \sum_{i}} \;
\varphi_{i} \right) \;\; , \eeq
where the contribution from the $i$-th object is $\varphi_{i} = {\mathcal S}
\times G(\mathbf{x}_{i}) \times \xi_{i}$.
%
%
That substructure produces as many positrons and antiprotons as if the entire
volume \beq \xi_{i} = {\displaystyle \int}_{i\text{-th \, mini-spike}} \left\{
{\displaystyle \frac{\delta \rho(\mathbf{x})}{\rho_{0}}} \right\}^{2} \, d^{3}
\mathbf{x} \label{xi_definition} \eeq were filled with the constant DM density
$\rho_{0}$. The mini-spike mass $M_{i}$ and \emph{intrinsic} boost $B_{i}$ are
related to the annihilation volume $\xi_{i}$ through
\beq \xi_{i} \equiv {\displaystyle \frac{B_{i} \, M_{i}}{\rho_{0}}} \;\; . \eeq
Note that the exotic flux $\varphi_{i}$ has no well-defined dependance on the
mass $M_{i}$ and the intrinsic boost $B_{i}$ if these quantities are considered
individually, since $\varphi_{i}$ only depends on the annihilation volume
$\xi_{i}$.

Should we know the exact location and annihilation volume of each mini-spike,
we would unambiguously derive the cosmic ray flux $\phi(E)$. This is not the
case, and for that matter we do not even know the number $N_{\rm BH}$ of these
objects. Very different distributions of the mini-spike population throughout
the galactic halo are possible. The positron and antiproton spectra at the
Earth are therefore affected by an uncertainty (which could be called a cosmic
variance) related to this lack of knowledge. A whole set of halo realizations
has to be considered, and specific semi-analytical tools have been developped
\cite{Lavalle:2006} to address this issue. The boost factor $B = \phi /
\phi_{s}$ \emph{at the Earth} is not unique and must be treated as a random
variable.
To do so, we have built a Monte-Carlo simulation nurtured by $\sim$ 200
different realizations of the mini-spike population which have been obtained in
Ref.~\cite{Bertone:2005xz} by evolving an initial distribution of IMBHs
orbiting in the Milky Way halo and by allowing the associated DM mini-halos to
be tidally destroyed during close encounters. The distribution of the number
$N_{\rm BH}$ of surviving mini-spikes is presented in \citefig{fig:number}. In
our Monte-Carlo simulation, $N_{\rm BH}$ is randomly drawn according to a
gaussian distribution. About 100 objects are expected within a 100 kpc
galactocentric radius and $\sim$60 of them populate the Galactic diffusive
halo.

%
\begin{figure}[h]
\centering
\includegraphics[width=\columnwidth]{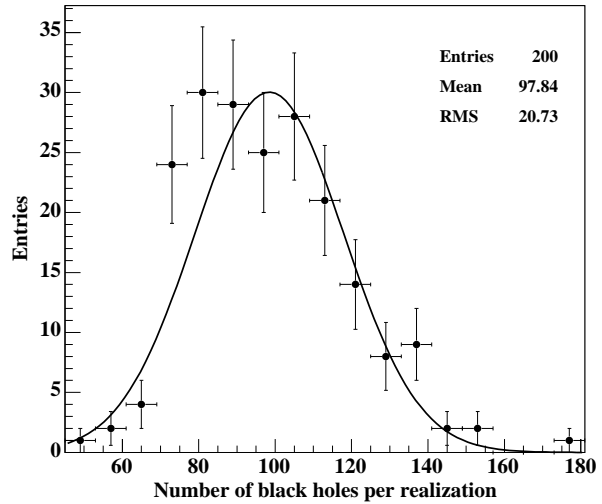}
\caption{ Distribution of the Monte-Carlo realizations of the galactic
mini-spike population -- extracted from Ref.~\cite{Bertone:2005xz} -- as a
function of the number $N_{\rm BH}$ of objects within a galactocentric radius
of 100 kpc.} \label{fig:number}
\end{figure}
%
\noindent If we set $N_{\rm BH}$ equal to its mean value, we may derive
analytically the average value and the variance of the boost factor $B$
according to the method presented in Ref.~\cite{Lavalle:2006}. We did not find
any significant correlation between the position $\mathbf{x}$ and annihilation
volume $\xi$ of the mini-spikes in the simulations of
Ref.~\cite{Bertone:2005xz}, so that they seem to be distributed independently
of the others. We readily infer an average boost factor of
\beq B_{\rm eff} = 1 \, + \, {\displaystyle \frac{\langle \phi_{r}
\rangle}{\phi_{s}}} = 1 \, + \, N_{\rm BH} \, {\displaystyle \frac{\langle \xi
\rangle \langle G \rangle}{\mathcal{I}}} \;\; . \label{B_eff} \eeq
The integral ${\mathcal{I}}$ stands for the convolution of the propagator $G$
with the smooth DM distribution $\rho_{s}$
\beq \mathcal{I} = {\displaystyle \int}_{\rm DM \, halo} \!\! G(\mathbf{x}) \,
\left\{ {\displaystyle \frac{\rho_{s}(\mathbf{x})}{\rho_{0}}} \right\}^{2} \,
d^{3} \mathbf{x} \;\; , \eeq
In the following, a NFW profile is assumed, where a value $\rho_\odot = 0.3
\;\text{GeV/cm}^3$ is considered for the Solar dark matter density. The
mini-spike annihilation volume and galactic position are respectively
distributed according to the probability functions $q(\xi)$ and $p(\mathbf{x})$
which are presented in Sections~\ref{subsec:inner_structure} and
\ref{subsec:position}. These probability laws allow to define the average
values
\beq \langle G^{n} \rangle = {\displaystyle \int}_{\rm DM \, halo} \! \left\{
G(\mathbf{x}) \right\}^{n} \, p(\mathbf{x}) \, d^{3} \mathbf{x} \;\; ,
\label{eq:Gn}\eeq
and
\beq \langle \xi^{n} \rangle = {\displaystyle \int_{0}^{+ \infty}} \, \xi^{n}
\, q(\xi) \; d\xi \;\; . \eeq
In practice, we will only be concerned with $n =$ 1 and 2. The boost scatter
$\sigma_{B}$ may be derived from the relation
\beq {\displaystyle \frac{\sigma_{B}}{B_{\rm eff}}} = {\displaystyle
\frac{{\sigma_{r}}/{\phi_{s}}}{1 + {\langle \phi_{r} \rangle}/{\phi_{s}}}}
\simeq {\displaystyle \frac{\sigma_{r}}{\langle \phi_{r} \rangle}} \;\; ,
\label{B_variance} \eeq
where the variance $\sigma_{r}$ of the random flux component $\phi_{r}$ is
given by
\beq \frac{\sigma_{r}^{2}}{\langle \phi_{r} \rangle^{2}} = \frac{1}{\langle
N_{\rm BH}\rangle}\left( \frac{\sigma_{\xi}^2}{\langle \xi \rangle^2} +
\frac{\sigma_{G}^2}{\langle G \rangle^2} +
\frac{\sigma_{\xi}^2\sigma_{G}^2}{\langle \xi \rangle^2\langle G \rangle^2}
\right)+\frac{\sigma_N^2}{\langle N_{\text{BH}}\rangle^2},\eeq
where $\sigma_{\xi}$, $\sigma_{G}$ and $\sigma_{N}$ stand for the variances of
the respective distributions. In Sections~\ref{sec:positrons} and
\ref{sec:antiprotons}, we will check the consistency of our Monte-Carlo results
in the light of the analytic expressions~(\ref{B_eff}) and (\ref{B_variance}).

\subsection{The inner structure of mini-spikes}
\label{subsec:inner_structure}

In the case where the DM profile {\it before} the formation of the IMBH follows
the commonly adopted NFW distribution~\cite{Navarro:1996he}, the final DM
density $\rho(r)$ around the IMBH will be described by a power law $r^{-7/3}$
in a region of size $R_\text{sp}$.
\begin{figure}[H]
\centering
\includegraphics[width=7cm]{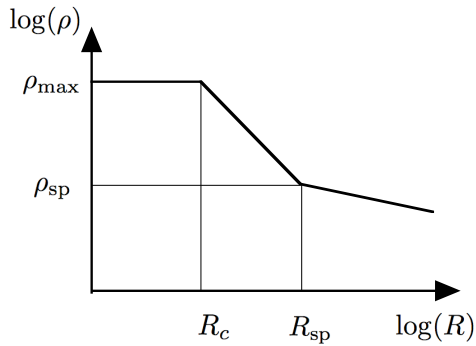}
\caption{Schematic representation of the inner structure of a
  mini-spike}
\label{fig:rho_r}
\end{figure}

At larger distances, the DM distribution has not been modified by the accretion
onto the IMBH and the density still falls down as $r^{-1}$. This envelope does
not contribute significantly to the DM annihilation and its associated
production of positrons and antiprotons.
On the contrary, the DM density steeply increases below $R_\text{sp}$ and
annihilations themselves set an upper limit to it
\begin{equation}
\rho_{\rm max} \approx {\displaystyle \frac{m_{\chi}}{\langle \sigma_{\rm ann}
v \rangle \, \tau}} \;\; ,
\end{equation}
where $\tau$ is the time elapsed since the formation of the mini-spike. We
denote by $R_{\rm c}$ the ``cut-off'' radius
below which the mini-spike core extends with uniform density $\rho_{\rm max}$.
For a typical value of $\tau = 10 \;\text{Gyr}$, we find
%
\begin{eqnarray}
\label{rho_max_num} \rho_{\rm max}=8.752\times10^{15}\times
\left(\frac{m_{\chi}}{{\rm GeV}}\right) \hspace{20ex}\\ \times\left(\frac{
\langle \sigma_\text{ann} v \rangle}{10^{-26}\,\text{cm}^3
\,\text{s}^{-1}}\right)^{-1}\;\text{M}_{\odot}\; \text{kpc}^{-3}\;.\nonumber
\end{eqnarray}

The relative extension of the mini-spike mantle ($R_{\rm c} \leq r \leq
R_\text{sp}$) with respect to the core ($r \leq R_{\rm c}$) is given by
\beq {\displaystyle \frac{R_\text{sp}}{R_{\rm c}}} = \left\{ \eta \equiv
{\displaystyle \frac{\rho_{\rm max}}{\rho_\text{sp}}} \right\}^{3/7} \;\; ,
\eeq
where $\rho_\text{sp} = \rho(R_\text{sp})$ is the DM density at the surface $r
= R_\text{sp}$ of the mantle. Integrating \citeeq{xi_definition}  over the
inner structure of the mini-spike leads to the annihilation volume
\beq \xi = {\displaystyle \frac{12}{5}} \, \pi \, R_\text{sp}^{3} \, \left\{
{\displaystyle \frac{\rho_\text{sp}}{\rho_{0}}} \right\}^{2} \, \left\{
{\displaystyle \frac{14}{9}} \, \eta^{5/7} \, - \, 1 \right\} \;\; .
\label{xi_mini_spike} \eeq
Due to the different mini-spike features (much less steep DM profile), the case
of scenario A leads to much lower annihilation volumes. This case is no more
mentioned in the following.
%
\begin{figure}[h]
\centering
\includegraphics[width=\columnwidth]{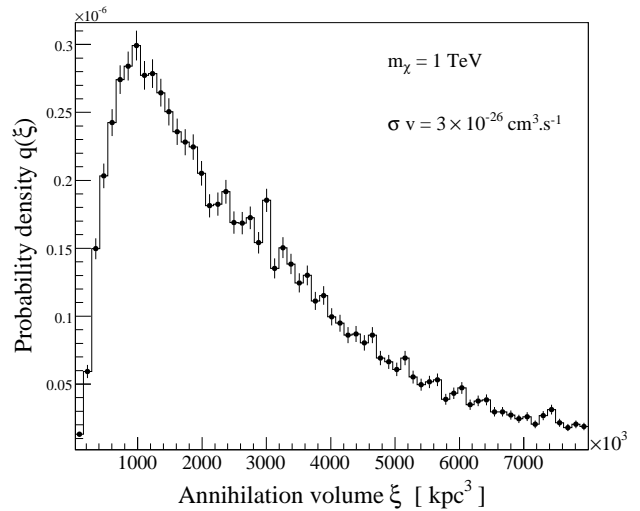}
\caption{The probability law $q(\xi)$ for the annihilation volume has been
derived from the Monte-Carlo results of Ref.~\cite{Bertone:2005xz}. We found no
correlation with the mini-spike position.} \label{fig:xi}
\end{figure}

\noindent We have derived the probability function for $\xi$ from the
Monte-Carlo results of Ref.~\cite{Bertone:2005xz} where the mini-spike radius
$R_\text{sp}$ and external density $\rho_\text{sp}$ are provided for each
object. This distribution -- featured in \citefig{fig:xi} -- is characterized
by very large values of the annihilation volume, with a tail extending up to
$\xi \sim 1.2 \times 10^{7}$ kpc$^{3}$.
On average, with a mini-spike radius $R_\text{sp} = 2.84$ pc and density
$\rho_\text{sp} = 48.51 \; {\rm M_{\odot} \, pc^{-3}}$, we infer a typical
annihilation volume $\xi \sim 3.3 \times 10^{6}$ kpc$^{3}$ for a benchmark
cross section $\langle \sigma_\text{ann} v \rangle= 3 \times 10^{-26} \,
\text{cm}^{3}\, \text{s}^{-1}$, a mass $m_\chi = 1\;\text{TeV}$ and $\rho_0
\equiv \rho_{\odot} = 0.3$ GeV cm$^{-3}$.
The distribution of $\xi$ happens to be log-normal. In our Monte-Carlo program,
we simulate a large number of different IMBH halo populations. For each
mini-spike, the annihilation volume $\xi$ is randomly chosen according to the
$q(\xi)$ distribution.

Notice finally that the individual mini-spike contribution $\varphi_{i}$ scales
as the product ${\mathcal S} \times \xi_{i}$ and is eventually proportional to
\beq \varphi_\text{mini-spike} \propto \langle \sigma_\text{ann} v
\rangle^{2/7} \; m_{\chi}^{-9/7} \;\; . \label{eq:scalelaw}\eeq
The cosmic ray signal depends weakly on the annihilation cross section since a
decrease of the cross section is partially compensated by a higher annihilation
volume. This issue is addressed with more details in
Section~\ref{sec:real_spectra}, where the influence of the injection spectrum
$g(E_S)$ will also be considered.

\begin{figure}[t]
\centering
\includegraphics[width=\columnwidth]{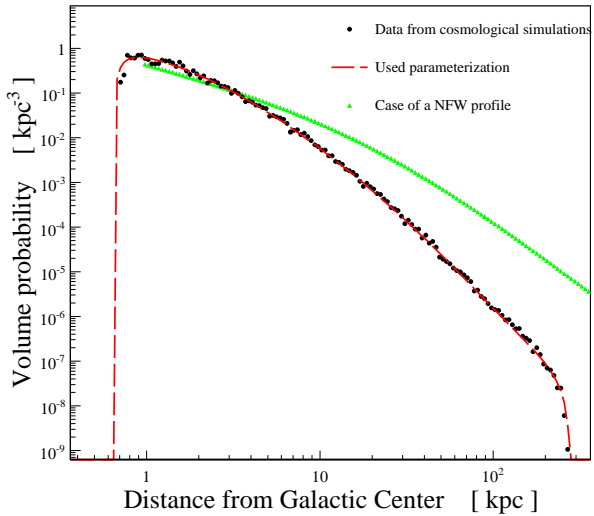}
\caption{Radial distribution of the mini-spikes, as extracted from the
numerical results of Ref.~\cite{Bertone:2005xz}.} \label{fig:radial_dist}
\end{figure}

\subsection{The galactic distribution of mini-spikes}
\label{subsec:position}

The Monte-Carlo simulation of IMBHs formation and evolution is used to derive
the distribution function $p(r)$ for the galactocentric radius of each
mini-spike that has survived tidal disruption. This function gives the
probability to find an object at galactocentric distance $r$ within a spherical
shell of thickness ${\rm d}r$. It is expressed in units of ${\rm kpc^{-3}}$ and
normalized to unity within the inner 100 kpc.
%
%
This volume distribution is presented in \citefig{fig:radial_dist} together
with the parameterized function used in the following to estimate the boost
factors.

The mini-spike number density drops quickly beyond a few hundred parsecs from
the galactic center. This is because encounters and mergers are much more
frequent in this region, and they are very efficient in disrupting dark matter
spikes. For those which have survived, the logarithmic slope of the radial
profile varies from $\sim 1.8$ in the inner region to $\sim 3.8$ outward. This
distribution is more peaked than for a typical NFW profile whose corresponding
volume probability is also drawn in \citefig{fig:radial_dist} for comparison.
%

\section{Results for positrons}
\label{sec:positrons}

This section is devoted to the positron signal. For pedagogical purposes, we
first consider a fiducial case in which dark matter consists in $m_\chi = 1
\;\text{TeV}$ particles fully annihilating at rest in electron-positron lines.
In this very simple frame, we scrutinize the positron flux enhancement that
mini-spikes produce with respect to a single smooth distribution of dark matter
in the Galaxy

The propagator for positrons is given by \citep{Lavalle:2006}. The energy at
which a positron is detected on Earth depends on its propagation history. It is
correlated to the distance over which it has diffused from the source. As a
result, the flux at a given (detected) energy depends on the spatial
distribution of the sources, i.e. on clumpiness. In this section, we study this
effect in a more quantitative way.

\subsection{The Monte-Carlo simulation}
\label{subsec:monte_carlo}

We have performed Monte-Carlo simulations of the mini-spike spatial
distribution in the halo. Each halo configuration yielded a value for the
positron flux  $\phi_{r} = \sum \varphi_i$, from which the average boost factor
and the corresponding variance could be computed. The probability density
functions for the number $N_{\rm BH}$ of IMBHs, their position $\mathbf{x}_{i}$
and annihilation volume $\xi_i$ were based on cosmological simulations from
Ref.~\cite{Bertone:2005xz}. The corresponding probability laws have been
presented in Section~\ref{sec:the_method}. The propagated flux is compared to
the flux produced by a smooth NFW profile to determine the boost factor.

\subsection{The numerical results}
\label{subsec:positron_results}

Monte-Carlo simulations involving $10^{6}$ realizations of the IMBHs population
have been performed. The mean value and the variance of the resulting boost
factors (for positron flux) at the Earth is sketched in \citefig{fig:B_E} as a
function of energy. The shaded area show the region where the boost factor is
expected to lie, with a 1-$\sigma$ and 2-$\sigma$ confidence level.
%

\begin{figure}[H] \centering
\includegraphics[width=\columnwidth]{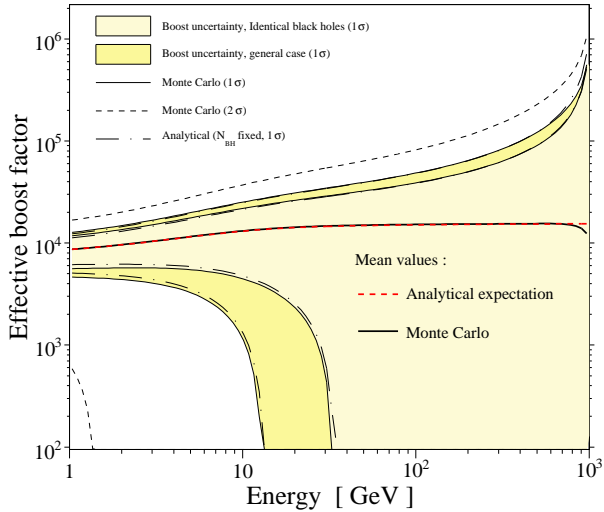}
\caption{Results from the Monte-Carlo simulations of the IMBHs population
inside the Milky Way are compared to the analytical computations of the
effective boost factor and its dispersion, for $m_{\chi} = 1 \; {\rm TeV}$.}
\label{fig:B_E}
\end{figure}

The yellow (grey) areas correspond to the 1 $\sigma$ region, the lighter one
being obtained by fixing the annihilation volumes $\xi_i$ to their mean
expected value, while the darker one corresponds to the general case for which
mini-spikes have values of $\xi_i$ drawn according to their true probability
distribution $q(\xi)$. In both cases, the dot-dashed curves stand for the 1
$\sigma$ contours obtained analytically. These curves are in good agreement
with the ones obtained from the Monte Carlo, the small increase of the variance
for the Monte Carlo with respect to the analytical expectation is due to the
variation of the black holes number from one Monte Carlo realization to
another, while this effect is not implemented in the analytical determination
of the boost factor. The fact that these curves are fairly close to each other
confirms that the dispersion of the number $N_{\rm BH}$ of IMBHs in the Milky
Way poorly influences the final dispersion of the boost. This figure also shows
that the boost factor can be very large, the expected value being of order 8000
(for a DM particle mass of 1 TeV).

\citefig{fig:B_E} shows that the variance increases with energy, and becomes
very large for $E \gtrsim 20 \;\text{GeV}$. It clearly appears that the boost
factor obeys two different statistical regimes. This is because the diffusive
range of positrons depends on energy and a positron emitted with energy $E_s$
looses energy as it propagates outwards from the source. The typical
propagation scale for a positron injected with energy $E_s$ and detected with
energy $E$ is given by
\beq\lambda_D \equiv \sqrt{4 K_0 \tau_E
\left(\epsilon^{\delta - 1}
  -\epsilon_{S}^{\delta-1} \right) /(1-\delta) } \; \;,\eeq
where $K_0$ and $\delta$ are respectively the normalization and the logarithmic
slope of the diffusion coefficient, and $\epsilon \equiv (E/1\;\rm{GeV})$. We
have considered the median parameters given in \cite{Donato:2003xg}, $K_0 =
0.0112\; \text{kpc}^2 \, \text{Gyr}^{-1}$, $L= 4 \;\text{kpc}$, $V_c = 12 \;
\text{km/s}$ and $\delta = 0.7$. The typical timescale for energy loss of 1 GeV
positrons is $\tau_E \approx 10^{16}\,\text{s}$. For a 1 TeV injected energy,
we infer a propagation length $\lambda_D \simeq 6.9 \; \text{kpc} \times
\sqrt{\epsilon^{-0.3}-0.12}$, which ranges from 0.1 kpc at a detected energy of
990 GeV to 6.4 kpc at 1 GeV.

At high energy, this diffusive range is very short, and positrons will be
detected only provided there happens to be a mini-spike very close to us. The
probability of this event is low at energy close to $E_s$, but the contribution
to the flux is high if it happens. The trade-off between these two trends leads
to a constant expectation value for the boost factor, but a large
shotnoise-like variance. Conversely, at low energy the diffusive range is
large, and the fluctuations in the number of sources contributing to the flux
becomes relatively small.

\begin{figure*}[t] \centering
\includegraphics[width=8.5cm]{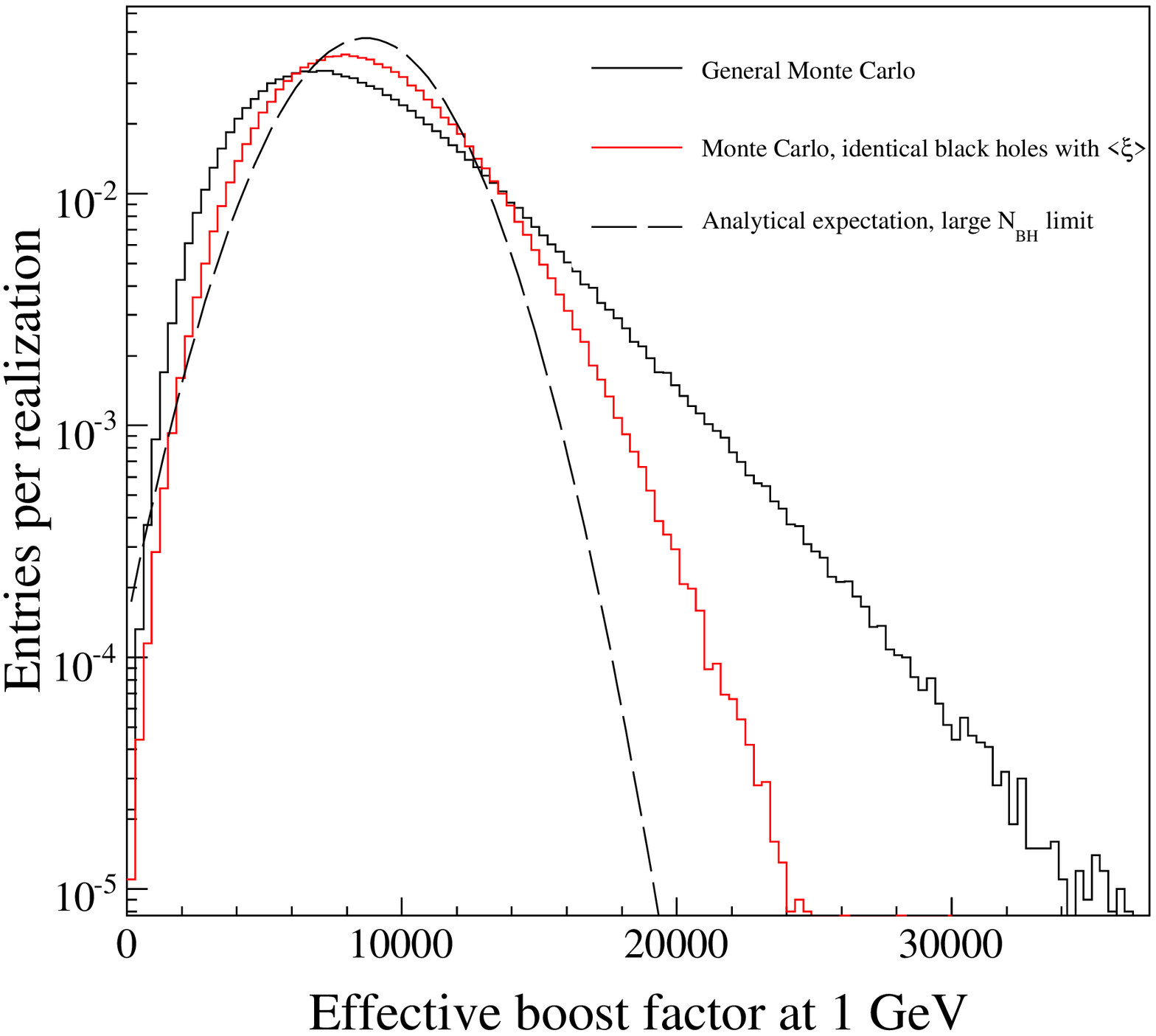}
\includegraphics[width=8.5cm]{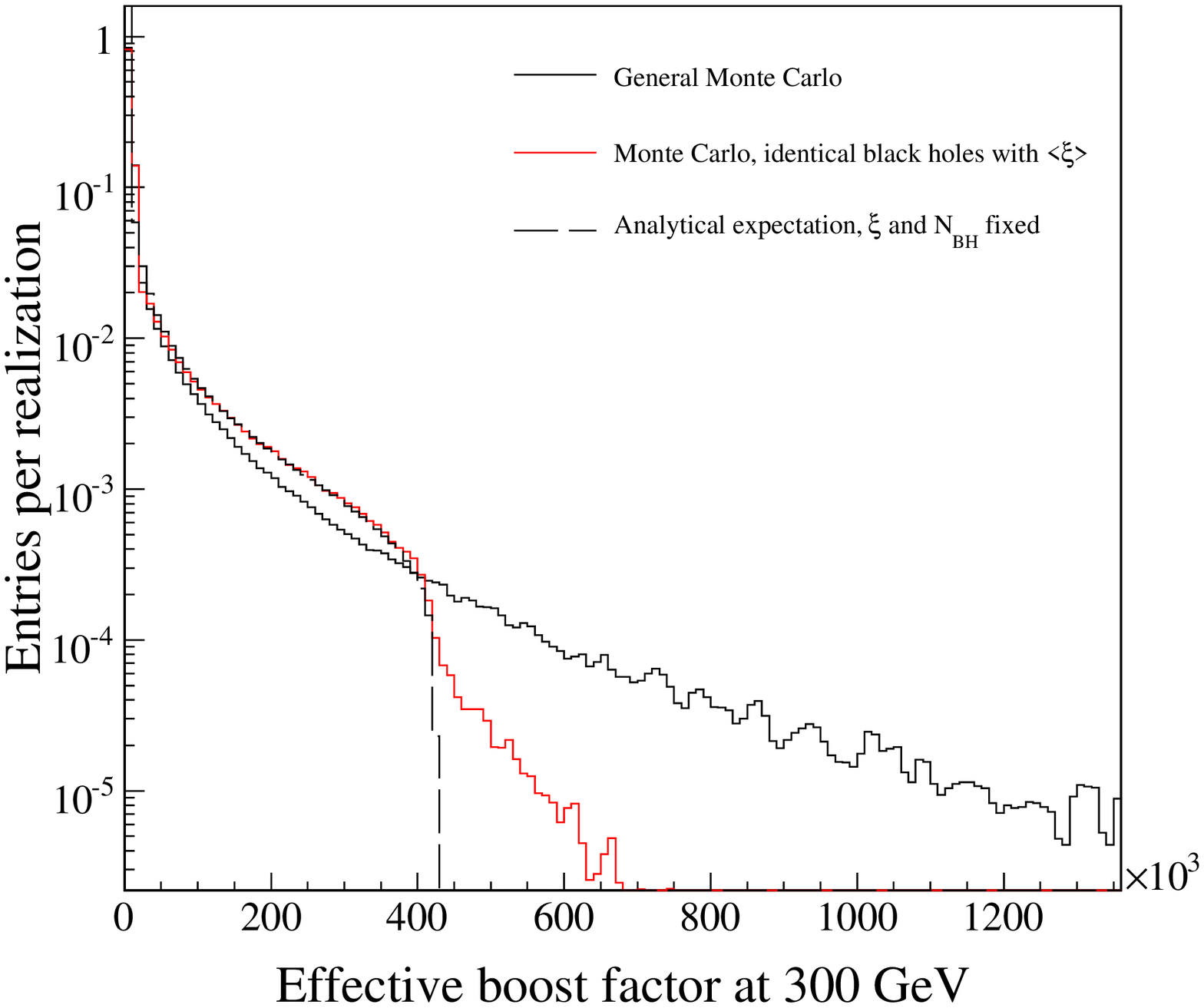}
\caption{Distribution of the boost factors at 1 GeV (left panel) and 300 GeV
(right panel) obtained with the Monte Carlo simulations and comparison to
analytical expectations (see text for further details). The gaussian
distribution discussed in the text is plotted in the left panel (long-dashed
curve). } \label{fig:B_Efix}
\end{figure*}

The distributions of the boost factor at 1 GeV obtained with Monte Carlo
simulations are displayed in the left panel of \citefig{fig:B_Efix}. The red
histogram corresponds to the case of identical mini-spikes with fixed
annihilation volume. It can be compared to the analytical estimate of the boost
factor distribution (dashed curve) in the limiting case where the number of
IMBHs is large enough for the central-limit theorem to be valid. If it were so,
the distribution of the boost factor would tend to be gaussian. One can see
that it is almost the case for identical objects. Though unrealistic, the case
in which ten times more IMBHs populate the Milky Way leads actually to a
gaussian distribution for $B(1\;{\rm GeV})$ (not shown here). The statistics
obtained in the general case is plotted in black, one can see that the effect
of having different annihilation volumes for different mini-spikes shifts the
distribution away from the gaussian behavior.

The right panel of \citefig{fig:B_Efix} displays the distribution of the boost
factor at 300 GeV. Again, the red histogram corresponds to the identical IMBHs
case and should be compared to the analytical estimate (dashed curve). The
latter includes only one IMBH inside the sensitivity volume so that the extra
events of the red curve correspond to the (very) rare situations in which two
mini-spikes contribute to the signal. This feature is not present in the
general case, for which the statistics is displayed in black~: it is erased by
the random choice of $\xi$.

\section{Results for antiprotons}
\label{sec:antiprotons}

\subsection{Propagator}
\label{subsec:antiproton_propagator}

The propagator of antiprotons has a larger range than for positrons, and the
effect of escape at the galactic boundaries must be considered. Modelling the
galaxy as a cylinder of radius $R\sim 20\,\text{kpc}$ and of unknown height
$2L$, in which antiprotons diffuse with an energy-dependent coefficient $K$,
the solution of the diffusion equation can be obtained as Bessel-Fourier
expansion over the $r$ and $z$ variables. It turns out that the side boundaries
($r=R$) can be neglected in most situations, as the $z=\pm L$ boundaries are
closer to the solar neighborhood and influence more strongly the measured
cosmic ray flux. As a consequence, the problem is essentially two-dimensional
and the propagator from any source to the Earth can be expressed as a fonction
or $r$ and $z$ only. Moreover, the sources (the mini-spikes) can be considered
as point-like as far as the propagation over galactic distances is considered,
and special care has to be taken for the singular $1/r$ dependance of $G(r,z)$.
This point has been exposed in detail in \cite{Bringmann:2006im}, along with
the corresponding analytical expressions for $G(r,z)$. The main difference with
the positron case is that the energy losses due to radiative losses are much
less important for antiprotons. If the diffusion is supposed to occur at
constant energy, then the spatial distribution of antiprotons created in a
given mini-spike is the same at every energy, and the boost factor should be
energy-independent (see below). There are actually at least two effects that
depend on energy, namely spallation and galactic wind. Their influence will be
discussed below, along with the results. We have considered the same median
parameters as in the positron case.

%
%
\subsection{The Monte-Carlo simulation}
\label{subsec:antiproton_montecarlo}

Following the same method as for positrons, we have performed Monte-Carlo
simulations of the IMBHs distribution, in order to estimate the expected
antiproton flux, along with the associated variance. The procedure is exactly
the same as before, and this section is devoted to the presentation of the
results.
%

\subsection{Expected values and variance of the boost factor}
\label{subsec:antiproton_boost}

The average boost factor and its scatter are computed, both by direct
evaluation of Eqs.~(\ref{B_eff}) and (\ref{B_variance}) as well as by the
Monte-Carlo simulation. Since energy losses are less important for antiprotons
than for positrons, we find that when spallation and galactic wind are
neglected, the boost factor does not depend on energy -- see
\citefig{fig:antiproton_boost}. When galactic wind is considered, propagation
becomes energy-dependent at low energy, and so does the boost factor -- see the
dotted line in \citefig{fig:antiproton_boost}. The uncertainty band widens
significantly below a few GeV. The importance of the galactic wind can be
estimated through the Peclet number defined as $\text{Pe}\equiv V_cL/K$, the
effect of the wind being negligible for small values of Pe. The diffusion
coefficient $K$ is a growing function of energy, so propagation is dominated by
diffusion at high energy, whereas the wind has a sizeable effect at low energy.
As the galactic wind is directed outward from the disk, it prevents antiprotons
from reaching the Earth. As a result, the range of diffusion is lowered and the
number of sources actually contributing to the signal is also lowered. This
explains the larger variance at low energy.

\begin{figure}[h]
 \includegraphics[width=\columnwidth]{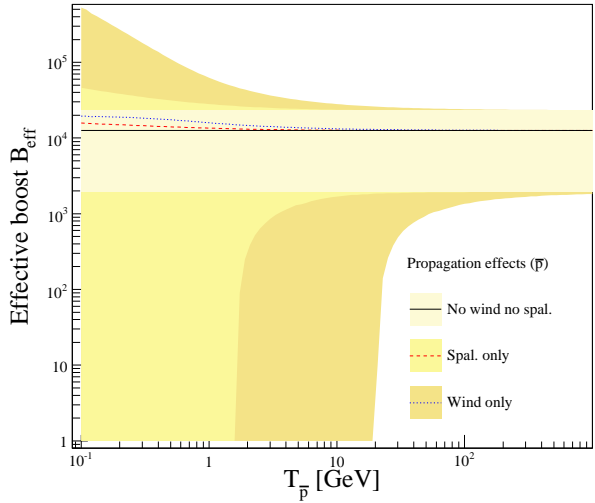}
    \caption{Expected value and variance of the boost factor of the
  antiproton signal as a function of kinetic energy, in the case of a  dark matter particle with
  mass $m_\chi = 1 \;\text{TeV}$ and for different $\rm\bar{p}$ propagation configurations.}
\label{fig:antiproton_boost}
\end{figure}

We see that the uncertainty on the antiproton flux is about one order of
magnitude in the best case, i.e. with no energy-dependent processes. When both
spallation and wind are considered, the prediction of the boost factor becomes
meaningless below $\sim$ 20 GeV. Notice however that when the thickness of the
diffusive halo is larger, more mini-spikes contribute to the signal at the
Earth and the variance decreases at all energies.


\section{A few realistic models for the DM particle}
\label{sec:real_spectra}

We have so far considered a generic case for the DM particle, with a mass of 1
TeV and a typical weak-scale annihilation cross section. This part is devoted
to the estimate of the positron and antiproton exotic fluxes within specific
particle physics models. We choose four typical DM particle candidates in the
frameworks of supersymmetry and extra-dimensions.

\begin{table*}[h]
\begin{tabular}{c|cc|cc|cc|cc}
&\multicolumn{2}{c|}{$\tilde{B}$} & \multicolumn{2}{c|}{$\tilde{H}$} & \multicolumn{2}{c|}{LZP} & \multicolumn{2}{c}{LKP} \\
\hline
\multirow{2}{*}{$m$} & \multicolumn{2}{c|}{\multirow{2}{*}{140 GeV}} & \multicolumn{2}{c|}{\multirow{2}{*}{108 GeV}} & \multicolumn{2}{c|}{\multirow{2}{*}{50 GeV}} &\multicolumn{2}{c}{ \multirow{2}{*}{1 TeV}} \\
 &&&&&&&&\\
\hline
\multirow{2}{*}{$\sigma v$ ($\times 10^{-26} \; {\rm cm^3\,s^{-1}}$)} & \multicolumn{2}{c|}{\multirow{2}{*}{0.26}} & \multicolumn{2}{c|}{\multirow{2}{*}{1.9}} & \multicolumn{2}{c|}{\multirow{2}{*}{2.04}} & \multicolumn{2}{c}{\multirow{2}{*}{1.7}}\\
&&&&&&&&\\
\hline
&&&&&&&&\\

\multirow{4}{*}{Final States}       & $b\bar{b}$ & 91\%     & $W^+W^-$ & 90\%   & $q\bar{q}$ & 74\%                     & $q\bar{q}_{up}$ & 11\% ($\times 3$)\\
                                    & $\tau^+\tau^-$ & 9\%  & $Z^0Z^0$ & 10\%   & $\nu\bar{\nu}$ & 17\%                 &  $q\bar{q}_{down}$ & 1\% ($\times 3$)\\
                                    &                &      &&                  & $\ell^+\ell^-$ &  2.88\% ($\times 3$) & $\nu\bar{\nu}$ & 4\% \\
                                    &&                      &&                  &&                                      & $\ell^+\ell^-$ & 20\% ($\times 3$)\\
&&&&&&&&\\
\hline \multirow{2}{*}{$\rho_{\rm max}$ ($\times 10^{17}\;{\rm
M_{\odot}\,kpc^{-3}}$)}& \multicolumn{2}{c|}{\multirow{2}{*}{47.1}} &
\multicolumn{2}{c|}{\multirow{2}{*}{4.97}} &
\multicolumn{2}{c|}{\multirow{2}{*}{2.15}} &
\multicolumn{2}{c}{\multirow{2}{*}{51.5}}\\
&&&&&&&&\\
\hline \multirow{2}{*}{$\langle \xi \rangle$ ($\times 10^{5}\;{\rm kpc^{3}}$)}&
\multicolumn{2}{c|}{\multirow{2}{*}{46.6}} &
\multicolumn{2}{c|}{\multirow{2}{*}{9.35}} &
\multicolumn{2}{c|}{\multirow{2}{*}{5.14}} &
\multicolumn{2}{c}{\multirow{2}{*}{49.6}}\\
\end{tabular}\vspace{.5cm}
\caption{Relevant parameters for the particle dark matter models
  discussed in Sec.~\ref{sec:real_spectra}.}
\label{tab:DMpart}
\end{table*}

All of these match the constraints from collider experiments and relic density.
The smooth halo is modelled by a NFW profile with a Solar dark matter density
$\rho_\odot = 0.3 \;\text{GeV/cm}^3$. In the previous computations (positronic
line, energy independent diffusion of antiprotons), the energy distribution at
the source $g(E_S)$ was not relevant since it cancelled in the ratio
$\phi_r/\phi_s$. In this section, we make predictions for the flux and we
perform the convolution of the propagator with the injection spectrum $g(E_S)$.
This is true in particular in the case of positrons, for which energy losses
are more significant than for antiprotons. The injection spectrum depends on
the final states of the DM particle annihilation process. In all the models we
considered, these spectra have been computed with
\texttt{PYTHIA}~\cite{pythia}. The annihilation cross section plays a
significant role as regards the ${\mathcal S}$ factor of \citeeq{Sfactor} as
well as the annihilation volume $\xi$ through the dependence of the maximum
density in the core, given by \citeeq{rho_max_num}.

\begin{figure*}[t!]
\includegraphics[width=8cm]{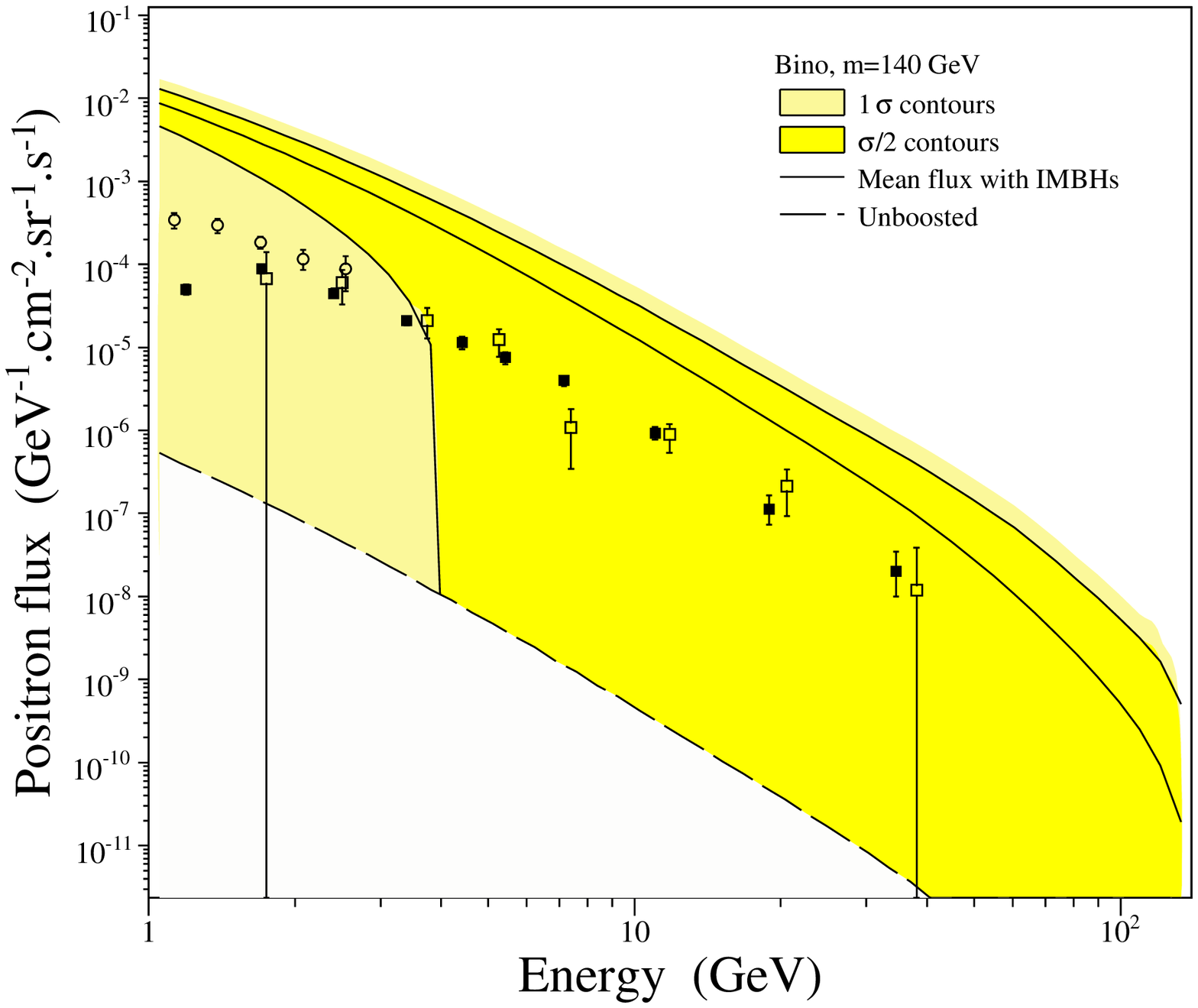}
\includegraphics[width=8cm]{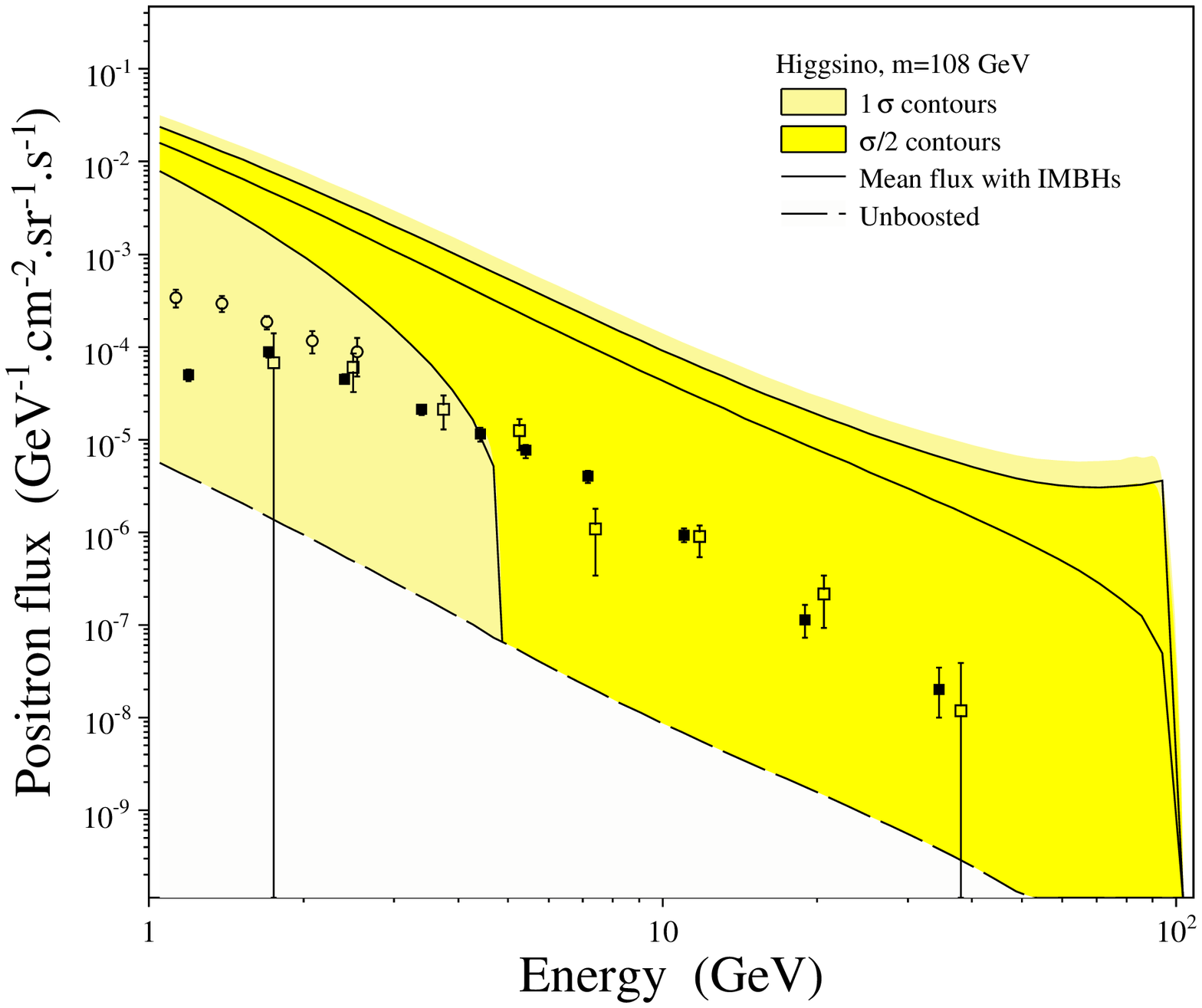}
\includegraphics[width=8cm]{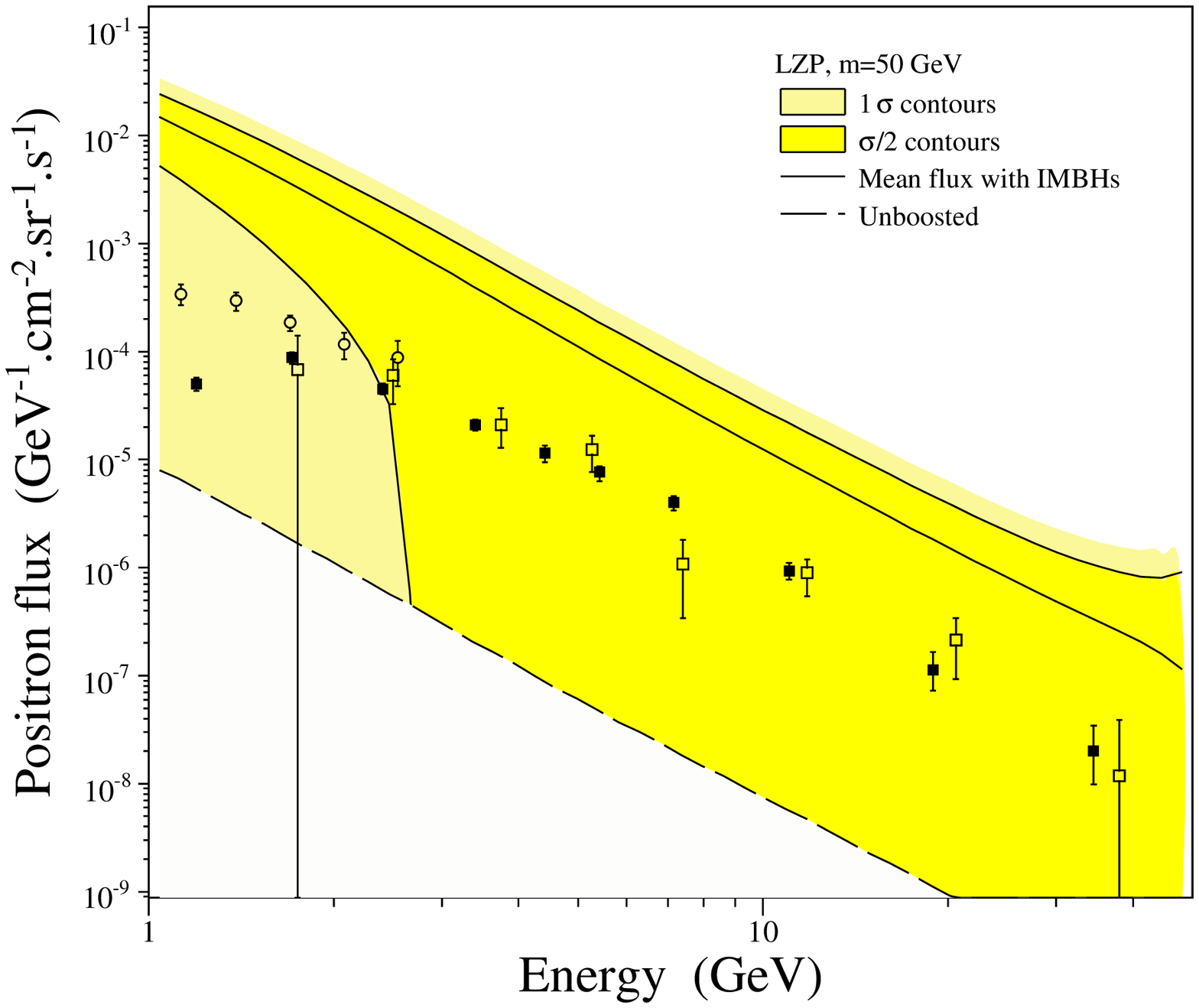}
\includegraphics[width=8cm]{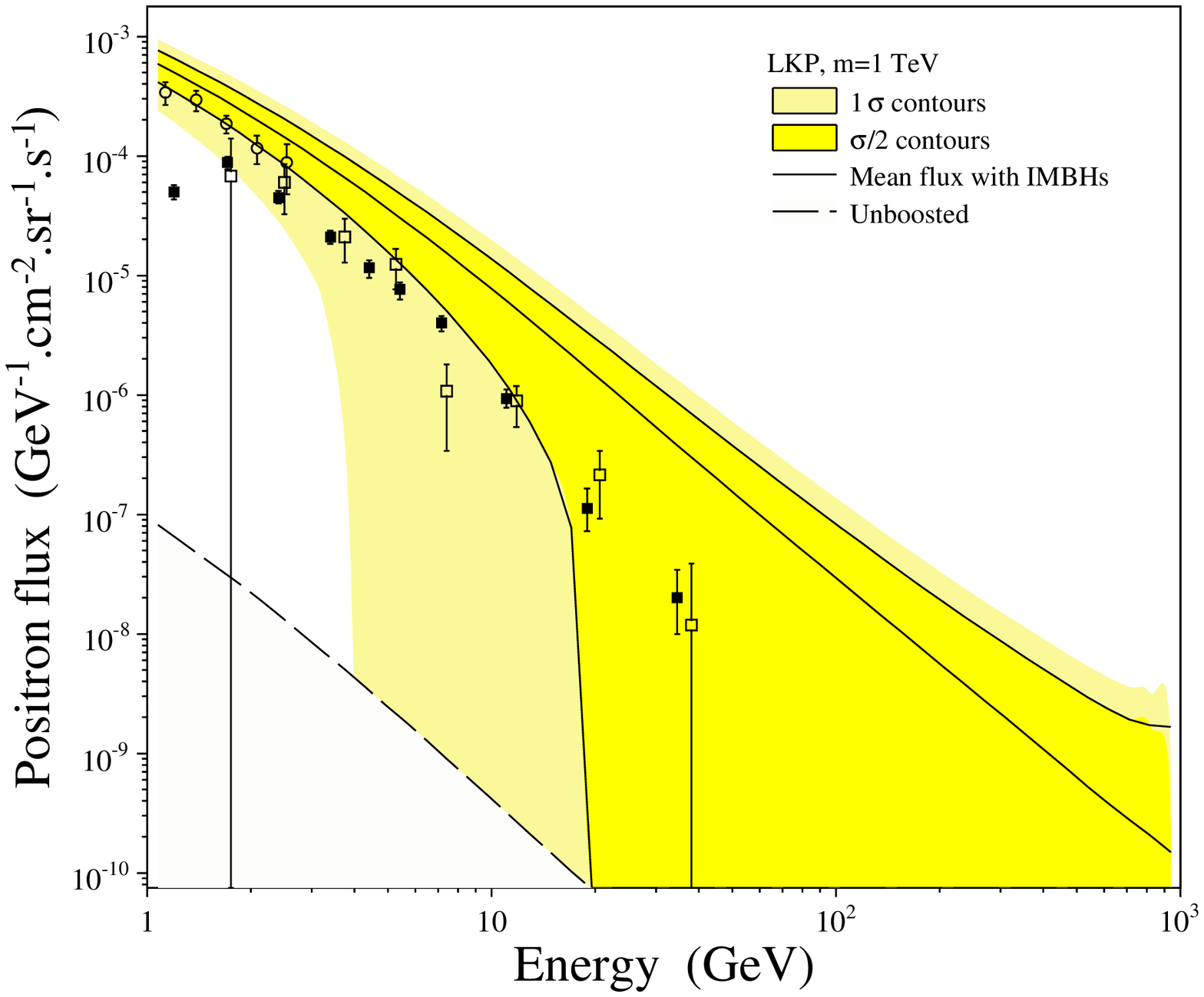}
\vspace{.5cm} \caption{Exotic positron fluxes in four particle physics models
(top : supersymmetric DM, bottom : Kaluza-Klein DM) within IMBHs mini-spike
scenario, and comparison to actual measurements. The bright and the pale yellow
areas respectively correspond to contours at the $\sigma/2$ and $1\;\sigma$
level.} \label{fig:realposit}
\end{figure*}
\begin{figure*}[t!]
\centerline{
  \includegraphics[width=\columnwidth]{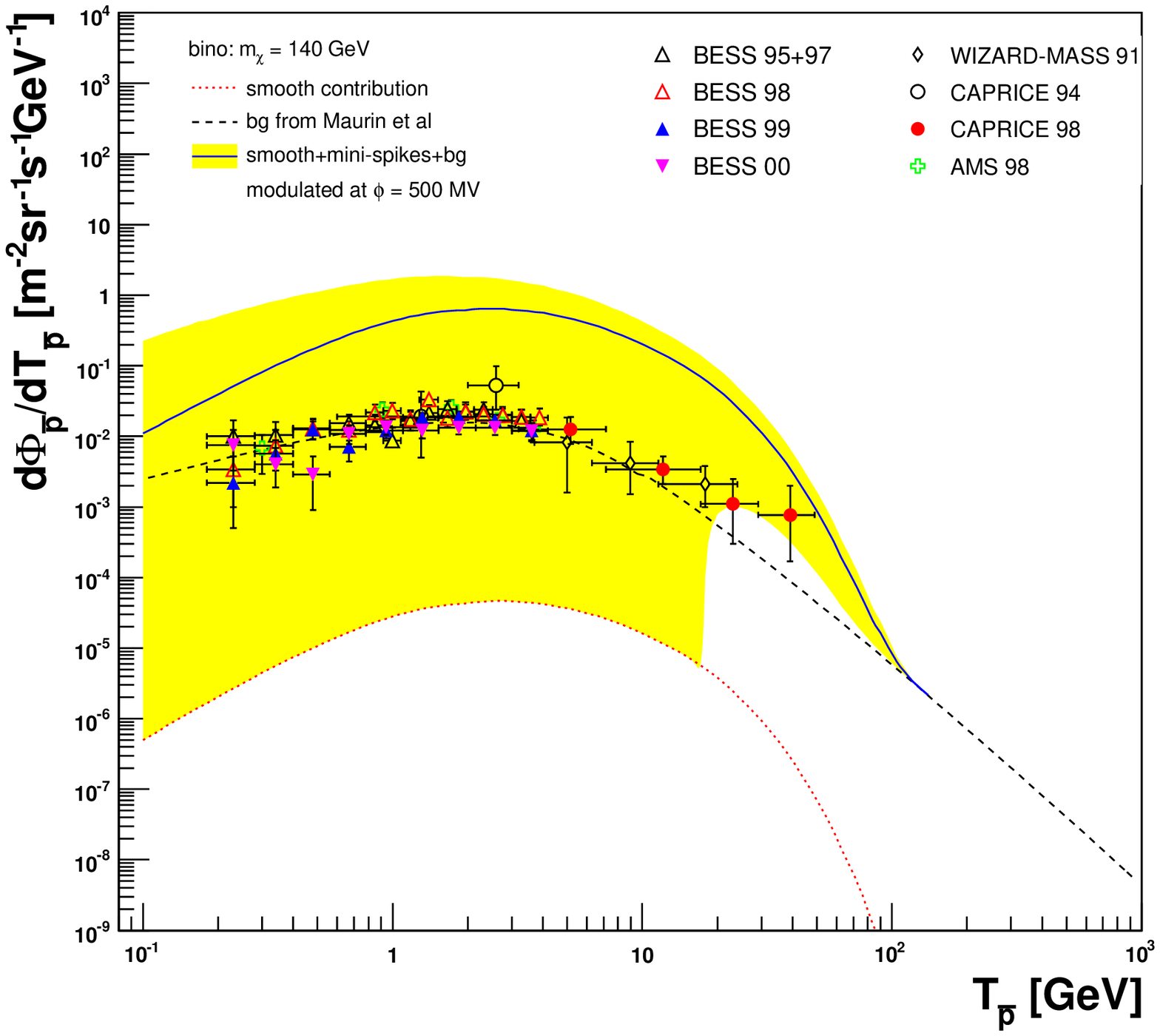}
  \includegraphics[width=\columnwidth]{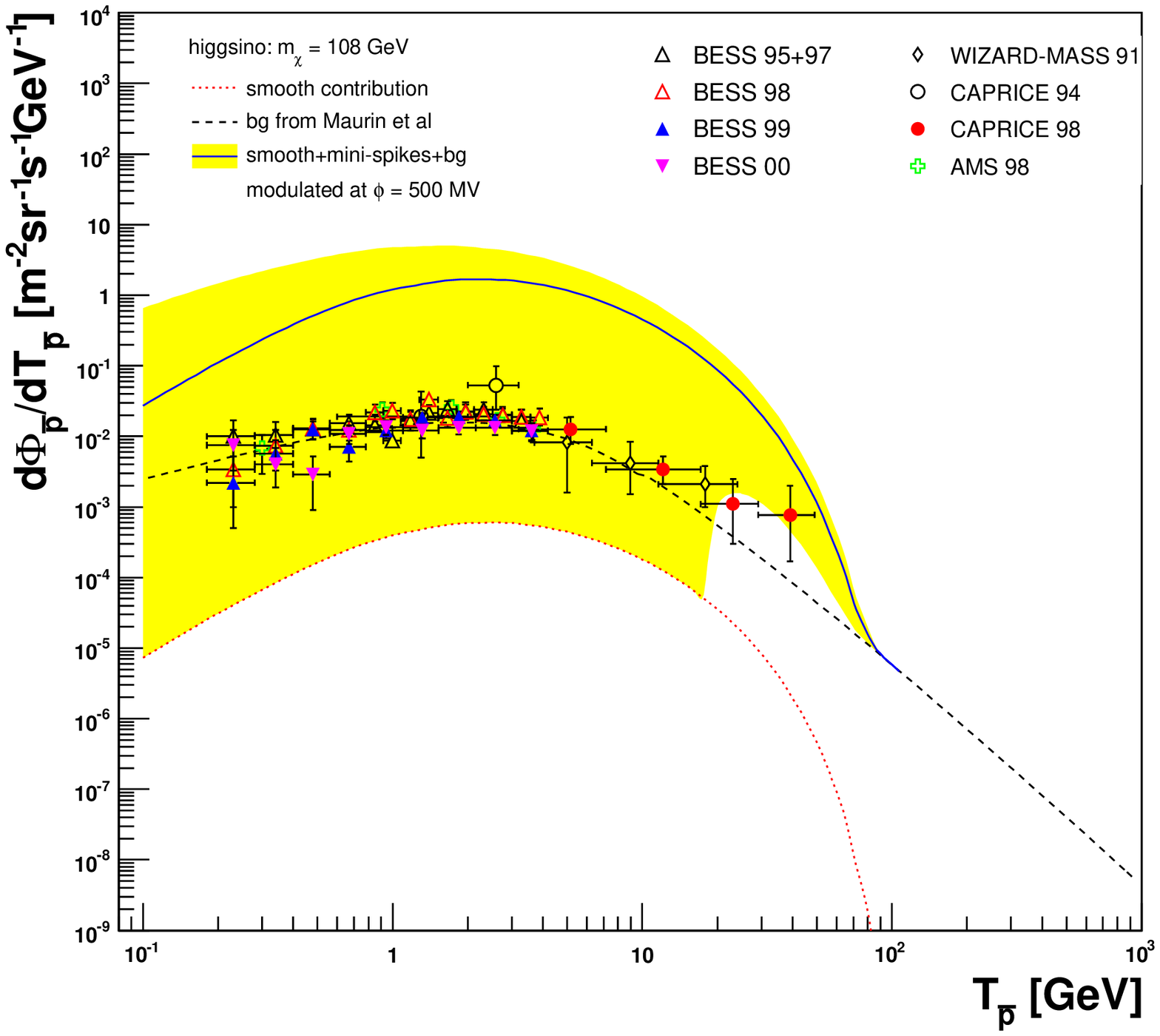}
} \centerline{
  \includegraphics[width=\columnwidth]{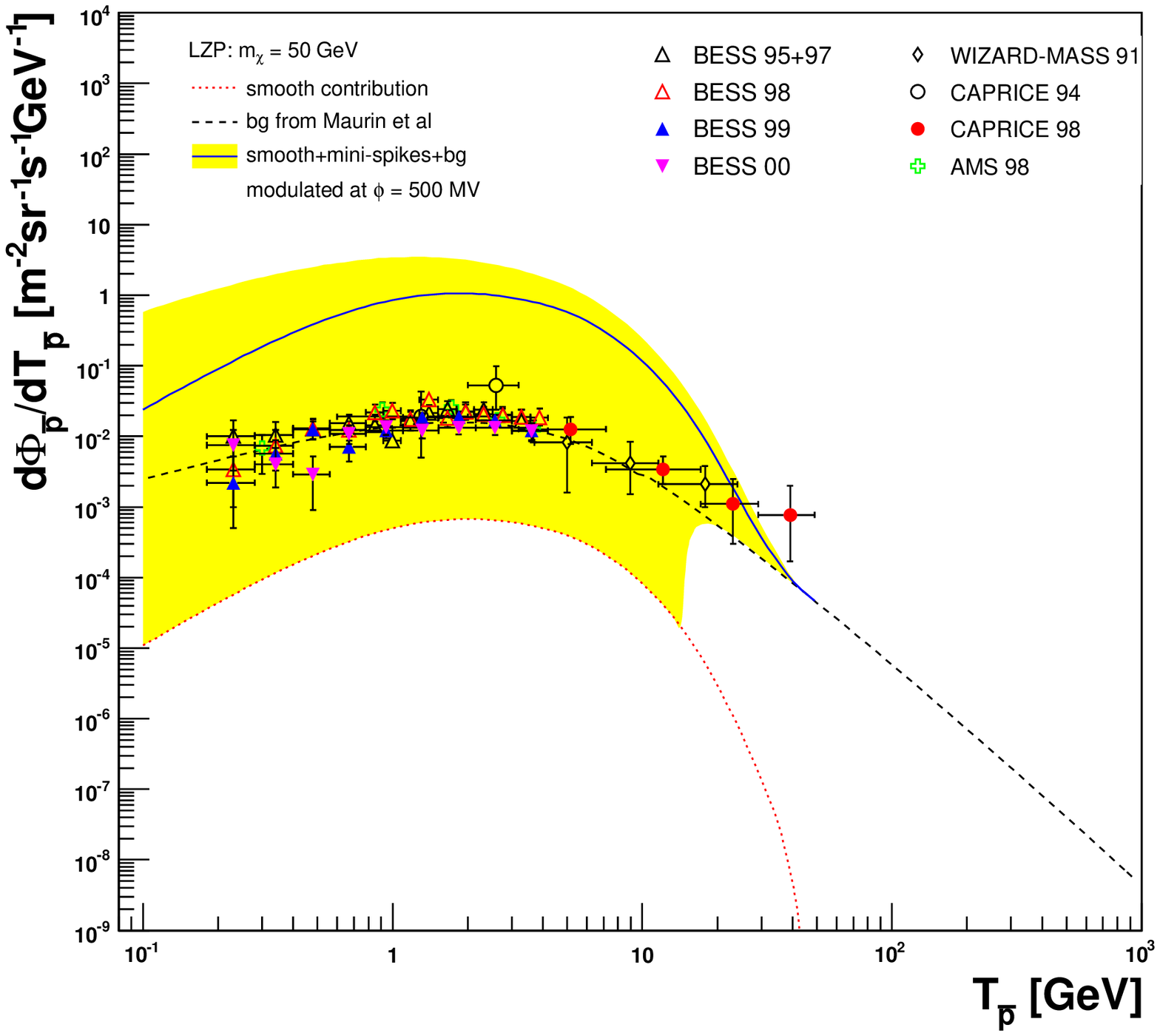}
  \includegraphics[width=\columnwidth]{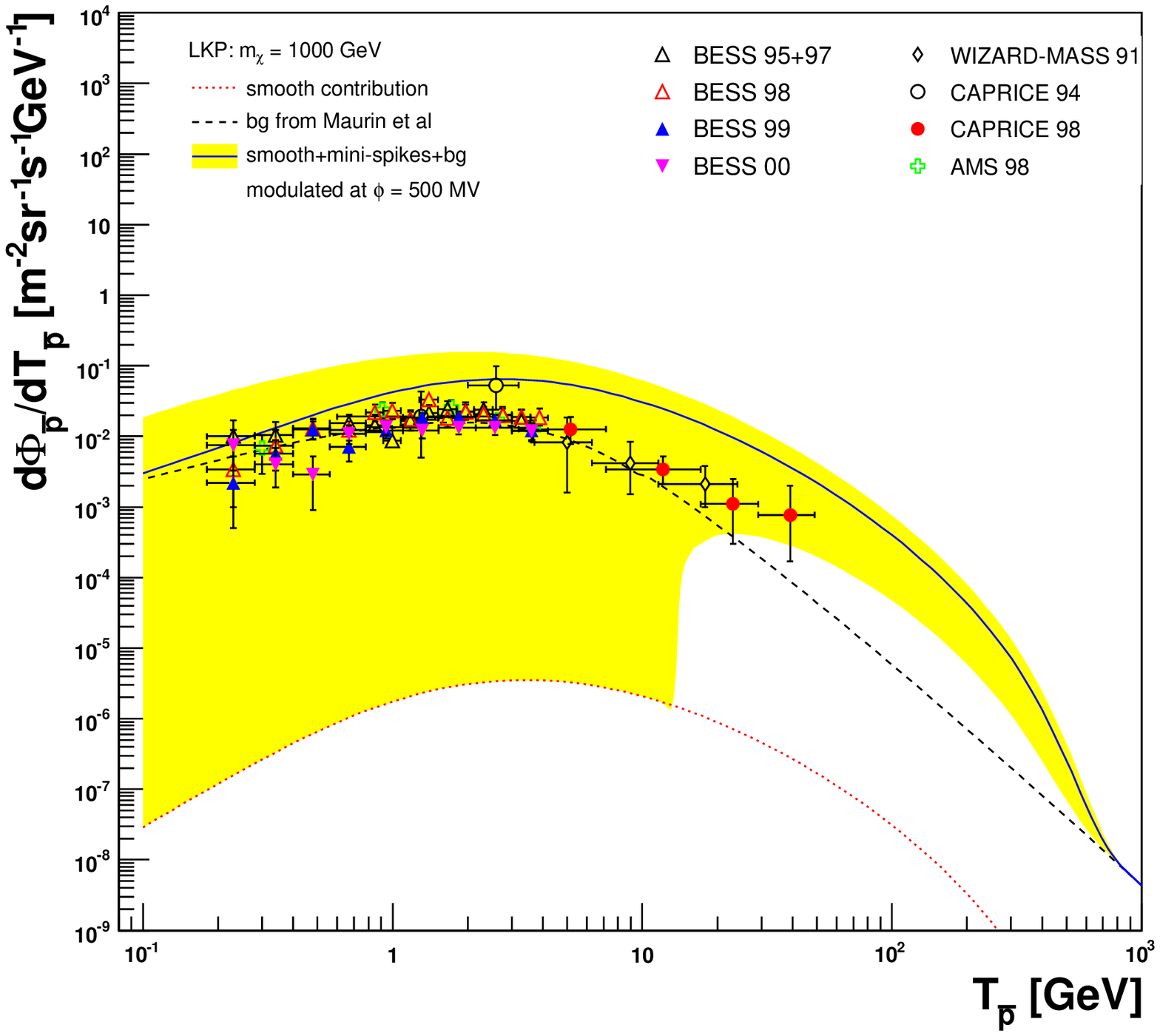}
} \caption{Antiproton fluxes as a function of kinetic energy in the case of a
bino dark matter particle with mass $m_\chi = 140 \;\text{GeV}$. The fluxes
were computed without (upper panel) or with (lower panel) wind (with a value
$V_c = 12\; \text{km/s}$), without (left) or with (right) spallation.}
\label{fig:realpbar}
\end{figure*}

In the framework of the Minimal Supersymmetric Standard Model (MSSM) with a
gravity driven supersymmetry breaking (mSUGRA), we choose two specific sets of
parameters leading to two typical DM candidates. The latter are a {\it bino}
and a {\it Higgsino} (respectively referred to as $\tilde{B}$ and $\tilde{H}$
in the following), depending on the field content of the lightest
supersymmetric particle. They differ mainly in their couplings to the Standard
Model particles and thus have different annihilation final states. The public
code \texttt{SuSpect}~\cite{suspect} is used to derive the weak-scale
parameters from the mSUGRA inputs by solving the renormalization group
equations. The cross sections for the different final states have been
determined with the \texttt{micrOMEGAs}~\cite{micro1,micro2} package.

The first Kaluza-Klein model considers warped extra-dimensions with a SO(10)
GUT, as described in~\cite{agashe}. In that context the DM particle is a right
handed Dirac neutrino. Its stability arises from the conservation of a
$\mathbf{Z}_{3}$ symmetry -- hence its name LZP for Lightest Z Particle.
To fulfill the relic density constrains, we choose $M_{KK}=6\,{\rm TeV}$ and
$m_{LZP}=50\,{\rm GeV}$ respectively for the Kaluza-Klein scale and the LZP
mass.
The second Kaluza-Klein model assumes universal extra-dimensions with
$R^{-1}=1\,{\rm TeV}$~\cite{servant}. Then, the DM particle is the lightest
particle with odd K-parity, which happens to be the first Kaluza-Klein
excitation of the $B^0$ hypercharge gauge boson, referred to as the $B^{(1)}$
(or LKP) with mass $m_{LKP}=1\,{\rm TeV}$.

All the relevant particle physics parameters and the inferred values regarding
the maximum density and the mean annihilation volume $\langle \xi \rangle$ are
summarized in Tab.~\ref{tab:DMpart}. Monte Carlo simulations are performed in
the context of these four models. The LZP being a Dirac particle, one has
$\delta=1/4$ in that case, whereas $\delta=1/2$ otherwise. The results are
presented in \citefig{fig:realposit} for positrons and in
\citefig{fig:realpbar} for antiprotons. The absolute fluxes are featured
together with the actual data points. The positron flux measurements are taken
from HEAT-$e^\pm$~\cite{heat1} and HEAT-pbar experiments~\cite{heat2} and two
independent analyses of AMS01 data~\cite{ams1,ams2}. For antiprotons, we
borrowed the results from \cite{bess1, bess2, caprice, ams98}.

We first discuss the positron case. In \citefig{fig:realposit}, the dashed line
is the unboosted, and therefore guaranteed, exotic flux in the framework of the
considered models. In each case, the yellow (grey) area represents the flux
uncertainty with IMBH mini-spikes. For clarity, the bright yellow contours
stand for the $\sigma/2$ level whereas the $1\;\sigma$ contours are indicated
by the pale yellow regions. In the four models we have considered, the positron
flux at 10 GeV is in the range between $\sim 6 \times 10^{-6}$ and $\sim 3
\times 10^{-5} \, \text{cm}^{-2} \, \text{s}^{-1} \, \text{GeV}^{-1} \,
\text{sr}^{-1}$. The exotic fluxes generated by IMBH mini-spikes are fairly
insensitive to the annihilation cross section, as a consequence of
\citeeq{eq:scalelaw}. We unexpectedly find that the flux is not very sensitive
to the mass. This is because the injection spectrum $g(E_S)$ softens the
$m_\chi^{-9/7}$ dependence of the annihilation rate inferred from that
relation. In addition, the variance of the flux is significantly smaller in the
LKP case. Indeed, the injection spectrum $g(E_S)$ extends up to 1 TeV and has
an important high energy contribution. As a consequence, the sensitivity sphere
below $\sim$10 GeV has a wider spread than for the three other DM candidates.
The LKP flux is less sensitive to the shotnoise associated with the random
realization of the IMBH population. The other cases feature a higher
dispersion, mainly due to the fact that the DM particle mass is lower and the
sensitivity sphere smaller.

We now discuss the antiproton case. The results are displayed in
\citefig{fig:realpbar}, along with the standard flux, for the same set of
propagation parameters as for the positrons (see Sec.~\ref{sec:positrons}). In
the case of a smooth halo, the exotic contribution is always lower than the
flux due to standard processes. When IMBH mini-spikes are considered, the
antiproton flux at 10 GeV is in the range between $\sim$3 $\times 10^{-2}$ and
$\sim$4 $\times 10^{-1} \, \text{m}^{-2} \, \text{s}^{-1} \, \text{GeV}^{-1} \,
\text{sr}^{-1}$ in the four models we have considered. We observe the same
delicate interplay between the mass dependence of the annihilation rate and the
energy behaviour of the injection spectrum $g(E_S)$ -- though the later is less
dependent on the final state species than for positrons. As before, the
antiproton flux is weakly dependent on the annihilation cross section. This is
particularly obvious in the two upper panels, between which $\langle
\sigma_\text{ann} v \rangle$ varies by a factor $\sim$7. Below $\sim$20 GeV,
the antiproton primary fluxes have a very large scatter, as explained in
Sec.~\ref{subsec:antiproton_boost}.

\section{Discussion and perspectives}
\label{sec:conclusion}

We have considered a scenario in which the formation of IMBHs generates
dramatic enhancements of the DM density in their vicinity. We have computed the
expected number of unmerged IMBHs in the Galactic halo, their spatial
distribution and the properties of their associated mini-spikes. We found high
values for the positron and antiproton boost factors, e.g. of order $10^4$ for
a fiducial DM particle with a 1 TeV mass. For an average mini-spike population,
the flux of primary positrons and antiprotons turns out to be one to two orders
of magnitude larger than the measured flux, and is fairly insensitive to the
specific properties of the DM candidate. In particular, the annihilation cross
section has little influence because of the presence of an annihilation plateau
in the cores of mini-spikes. This interesting feature allows configurations in
the particle physics parameter space which are usually disregarded -- because
of their very low cross sections -- to become testable. As the average flux
signal exceeds the data points, we could naively conclude that the mini-spike
scenario is already ruled out by observation. However, because of the small
number of objects, the variance associated to the positron and antiproton
signal is enormous and no definitive conclusion can be reached. We leave for
further investigation the quantitative estimate of the confidence level at
which the the mini-spike scenario agrees with measurements.

Notice finally that this variance is small at low energy for positrons, and on
the contrary at high energy for antiprotons. Should a cut-off above some
specific energy $E_c$ be detected in the positron flux, the natural
interpretation would lead to a DM particle mass $m_\chi \sim E_c$. The possible
existence of substructures leaves room to more subtle explanations. In
particular, if an excess in the antiproton flux were found above $E_c$, we
would conclude that we live in a particular Milky Way realization where
substructures are far from the Earth. The farther those substructures, the
smaller $E_c$ with respect to $m_\chi$, hence the possibility of a
misinterpretation of the data. This illustrates the fact that combining various
indirect signals could give information about the spatial distribution of DM.

\begin{acknowledgments}
We thank Andrew Zentner for early collaboration and for making available the
numerical realizations of IMBHs in the Milky Way, on which our calculations are
based. We thank the French Programme National de Cosmologie (PNC) for its
financial support. P.B. wishes to thank St\'ephane Ranchon and the H.E.S.S.
team of LAPP for providing help and CPU for this project. G.B. acknowledges the
support, during the first part of this project, by the Helmholtz Association of
National Research Centres. J.L. is grateful to the French GDR SUSY and the
CPPM-ANTARES group for having supported this work. We also thank D. Hooper for
valuable comments.
\end{acknowledgments}

\newpage



\begin{thebibliography}{99}

\bibitem{Bergstrom:2000pn}
  L.~Bergstrom,
  Rept.\ Prog.\ Phys.\  {\bf 63} (2000) 793
  [arXiv:hep-ph/0002126].

\bibitem{Munoz:2003gx}
  C.~Munoz,
  Int.\ J.\ Mod.\ Phys.\ A {\bf 19}, 3093 (2004)
  [arXiv:hep-ph/0309346].

\bibitem{Bertone:2004pz}
  G.~Bertone, D.~Hooper and J.~Silk,
  Phys.\ Rept.\  {\bf 405} (2005) 279
  [arXiv:hep-ph/0404175].

\bibitem{glast} http://www-glast.stanford.edu/

\bibitem{canga} http://icrhp9.icrr.u-tokyo.ac.jp/index.html

\bibitem{hess} http://www.mpi-hd.mpg.de/hfm/HESS/HESS.html

\bibitem{magic} http://hegra1.mppmu.mpg.de/MAGICWeb/

\bibitem{veritas} http://veritas.sao.arizona.edu/index.html

\bibitem{icecube}
  J.~Ahrens {\it et al.}  [The IceCube Collaboration],
  Nucl.\ Phys.\ Proc.\ Suppl.\  {\bf 118}, 388 (2003)
  [arXiv:astro-ph/0209556].

\bibitem{Aslanides:1999vq}
  E.~Aslanides {\it et al.}  [ANTARES Collaboration],
  arXiv:astro-ph/9907432.

\bibitem{positrons1}
  G.~L.~Kane, L.~T.~Wang and J.~D.~Wells,
  Phys.\ Rev.\ D {\bf 65}, 057701 (2002).

\bibitem{positrons2}
  M.~Kamionkowski and M.~S.~Turner,
  Phys.\ Rev.\ D {\bf 43}, 1774 (1991).

\bibitem{positrons3}
  M.~S.~Turner and F.~Wilczek,
  Phys.\ Rev.\ D, {\bf 42}, 1001 (1990).

\bibitem{positrons4}
  A.~J.~Tylka,
  Phys.\ Rev.\ Lett., {\bf 63}, 840 (1989).

\bibitem{positrons5}
  G.~L.~Kane, L.~T.~Wang and T.~T.~Wang,
  Phys.\ Lett.\ B {\bf 536}, 263 (2002).

\bibitem{positrons6}
  E.~A.~Baltz and J.~Edsjo,
  Phys.\ Rev.\ D {\bf 59} (1999) 023511
  [arXiv:astro-ph/9808243].

\bibitem{Baltz:2001ir}
  E.~A.~Baltz, J.~Edsjo, K.~Freese and P.~Gondolo,
  Phys.\ Rev.\ D {\bf 65} (2002) 063511
  [arXiv:astro-ph/0109318].

\bibitem{Cheng:2002ej}
  H.~C.~Cheng, J.~L.~Feng and K.~T.~Matchev,
  Phys.\ Rev.\ Lett.\  {\bf 89}, 211301 (2002)
  [arXiv:hep-ph/0207125].

\bibitem{Lavalle:2006}
  J.~Lavalle, J.~Pochon, P.~Salati and R.~Taillet,
  Astron. Astrophys. \textbf{462}, 827 (2007)
  \\\ [arXiv:astro-ph/0603796].

\bibitem{antiproton1}
  A.~Bottino, F.~Donato, N.~Fornengo and P.~Salati,
  Phys.\ Rev.\ D {\bf 58}, 123503 (1998).

\bibitem{Donato:2003xg}
  F.~Donato, N.~Fornengo, D.~Maurin, P.~Salati and R.~Taillet,
  Phys. Rev. D \textbf{69}, 063501 (2004)
  \\\ [arXiv:astro-ph/0306207].

\bibitem{Bergstrom:1999jc}
  L.~Bergstrom, J.~Edsjo and P.~Ullio,
  ApJ \textbf{526}, 215 (1999)
  \\\ [arXiv:astro-ph/9902012].

\bibitem{Lionetto}
 A.M. Lionetto, A. Morselli, V. Zdravkovic, JCAP {\bf 09}, 010 (2005) \\\ [astro-ph/0502406]


\bibitem{Picozza:2006nm}
  P.~Picozza {\it et al.},
  arXiv:astro-ph/0608697.

\bibitem{ams02} http://ams.cern.ch/

\bibitem{Bertone:2006nq}
  G.~Bertone,
  Phys.\ Rev.\ D {\bf 73} (2006) 103519
  [arXiv:astro-ph/0603148].

\bibitem{Miller:2003sc}
  M.~C.~Miller and E.~J.~M.~Colbert,
  Int.\ J.\ Mod.\ Phys.\ D {\bf 13} (2004) 1
  [arXiv:astro-ph/0308402].

\bibitem{Fryer:2001}
  Fryer, C.~L., \& Kalogera, V.\ 2001,  Astrophys.\ J.\, 554, 548

\bibitem{Narayan:2003fy}
  R.~Narayan,
  arXiv:astro-ph/0310692.

\bibitem{Ferrarese:2005}
  Ferrarese, L., \& Ford, H.\ 2005, Space Science Reviews, 116, 523

\bibitem{kormendy:1995}
  Kormendy, J., \& Richstone, D.\ 1995,  Ann.\ Rev.\ Astron.\ \& Astrophys., 33, 581

\bibitem{Ferrarese:2000se}
  L.~Ferrarese and D.~Merritt,
  Astrophys.\ J.\  {\bf 539} (2000) L9
  [arXiv:astro-ph/0006053].

\bibitem{McLure:2001uf}
  R.~J.~McLure and J.~S.~Dunlop,
  Mon.\ Not.\ Roy.\ Astron.\ Soc.\  {\bf 331} (2002) 795
  [arXiv:astro-ph/0108417].

\bibitem{Gebhardt:2000fk}
  K.~Gebhardt {\it et al.},
  Astrophys.\ J.\  {\bf 539} (2000) L13
  [arXiv:astro-ph/0006289].

\bibitem{Tremaine:2002js}
  S.~Tremaine {\it et al.},
  Astrophys.\ J.\  {\bf 574} (2002) 740
  [arXiv:astro-ph/0203468].

\bibitem{Fan:2001ff}
  X.~Fan {\it et al.}  [SDSS Collaboration],
  Astron.\ J.\  {\bf 122} (2001) 2833
  [arXiv:astro-ph/0108063].

\bibitem{barth:2003}
  Barth, A.~J., Martini, P., Nelson, C.~H., \& Ho, L.~C.\ 2003, Astrophys. Lett.\, 594, L95.

\bibitem{Willott:2003xf}
  C.~J.~Willott, R.~J.~McLure and M.~J.~Jarvis,
  Astrophys.\ J.\  {\bf 587} (2003) L15
  [arXiv:astro-ph/0303062].

\bibitem{haiman:2001}
  Haiman, Z., \& Loeb, A.\ 2001, Astrophys.\ J.\ , 552, 459.

\bibitem{Bertone:2005xz}
  G.~Bertone, A.~R.~Zentner and J.~Silk,
  Phys.\ Rev.\ D {\bf 72} (2005) 103517 [arXiv:astro-ph/0509565].

\bibitem{Madau:2001}
  Madau, P., \& Rees, M.~J.\ 2001, Astrophys.\ J.\ Lett.\, 551, L27.

\bibitem{Zhao:2005zr}
  H.~S.~Zhao and J.~Silk,
  arXiv:astro-ph/0501625.

\bibitem{islamc:2004}
  R.~Islam, J.~Taylor and J.~Silk,
  Mon.\ Not.\ Roy.\ Astron.\ Soc.\  {\bf 354} (2003) 443.

\bibitem{islamb:2004}
  R.~Islam, J.~Taylor and J.~Silk,
  Mon.\ Not.\ Roy.\ Astron.\ Soc.\  {\bf 354} (2004) 427.

\bibitem{Koushiappas:2003zn}
  S.~M.~Koushiappas, J.~S.~Bullock and A.~Dekel,
  Mon.\ Not.\ Roy.\ Astron.\ Soc.\  {\bf 354} (2004) 292
  [arXiv:astro-ph/0311487].

\bibitem{peebles:1972}
  Peebles, P.~J.~E.\ 1972, Astrophys.\ J.\, 178, 371.

\bibitem{young:1980}
  P.~Young, 1980, Astrophys.\ J.\ {\bf 242} (1980), 1232.

\bibitem{Ipser:1987ru}
  J.~R.~Ipser and P.~Sikivie,
  Phys.\ Rev.\ D {\bf 35} (1987) 3695.

\bibitem{Quinlan:1995}
  Quinlan, G.~D., Hernquist, L., \& Sigurdsson, S.\ 1995,
  Astrophys.\ J.\  {\bf 440}, 554.

\bibitem{Gondolo:1999ef}
  P.~Gondolo and J.~Silk,
  Phys.\ Rev.\ Lett.\  {\bf 83} (1999) 1719,
  [arXiv:astro-ph/9906391].

\bibitem{Merritt:2003qc}
  D.~Merritt,
  Proceedings of Carnegie Observatories Centennial Symposium
  {\it Coevolution of Black Holes and Galaxies}
  [arXiv:astro-ph/0301257].

\bibitem{Bertone:2005xv}
  G.~Bertone and D.~Merritt,
  Mod.\ Phys.\ Lett.\ A {\bf 20} (2005) 1021
  [arXiv:astro-ph/0504422].

\bibitem{Bertone:2005hw}
  G.~Bertone and D.~Merritt,
  Phys.\ Rev.\ D {\bf 72} (2005) 103502
  [arXiv:astro-ph/0501555].

\bibitem{Ullio:2001fb}
  P.~Ullio, H.~Zhao and M.~Kamionkowski,
  Phys.\ Rev.\ D {\bf 64}, 043504 (2001)
  [arXiv:astro-ph/0101481].

\bibitem{Merritt:2002vj}
  D.~Merritt, M.~Milosavljevic, L.~Verde and R.~Jimenez,
  arXiv:astro-ph/0201376.

\bibitem{Merritt:2003eu}
  D.~Merritt,
  arXiv:astro-ph/0301365.

\bibitem{Merritt:2006mt}
  D.~Merritt, S.~Harfst and G.~Bertone,
  arXiv:astro-ph/0610425.

\bibitem{Navarro:1996he}
  J.~F.~Navarro, C.~S.~Frenk and S.~D.~M.~White,
  Astrophys.\ J.\ {\bf 490}, (1997) 493.

\bibitem{Navarro:04b}
  J. F. Navarro {\it et al.},
  {\it Mon. Not. R. Astron. Soc.} {\bf 349}, 1039 (2004).

\bibitem{Reed:05}
  D.~Reed {\it et al.},
  {\it Mon. Not. R. Astron. Soc.} {\bf 357}, 82 (2005).

\bibitem{Merritt:05}
  D.~Merritt, J.~Navarro, A.~Ludlow, and A.~Jenkins,
  astro-ph/0502515 (2005).

\bibitem{Koushiappas:2005qz}
  S.~M.~Koushiappas and A.~R.~Zentner,
  Astrophys.\ J.\ {\bf 639}, (2006) 7.

\bibitem{appel}
  T.~Appelquist, H.~C.~Cheng and B.~A.~Dobrescu,
  Phys.\ Rev.\ D {\bf 64}, (2001) {035002}.

\bibitem{G_pbar_0609522}
  D.~Maurin, R.~Taillet and C.~Combet,
  [arXiv:astro-ph/0609522].

\bibitem{Bringmann:2006im}
  T.~Bringmann and P.~Salati,
  [arXiv:astro-ph/0612514].

\bibitem{suspect} A. Djouadi, J.L. Kneur, G. Moultaka, G., 2002, \\\ [arXiv:hep-ph/0211331]\\\ http://www.lpta.univ-montp2.fr/~kneur/Suspect
\bibitem{pythia} T. Sj\"ostrand, S. Mrenna, Peter Skands, 2006, {\it JHEP}, {\bf 0605}, 026 \\\ [arXiv:hep-ph/0603175]
\bibitem{micro1} G. B\'elanger, F. Boudjema, A. Pukhov, A. Semenov, 2002, {\it Comput. Phys. Commun.}, {\bf 149}, 103   \\\ [arXiv: hep-ph/0112278]
\bibitem{micro2} G. B\'elanger, F. Boudjema, A. Pukhov, A. Semenov, 2006, {\it Comput. Phys. Commun.}, {\bf 174}, 577  \\\ [arXiv: hep-ph/0405253]
\bibitem{agashe} K. Agashe, G. Servant, 2004, {\it Phys. Rev. Lett.}, {\bf 93}, 231805  \\\ [arXiv: hep-ph/0403143]
\bibitem{servant} G. Servant, T. Tait, 2003, {\it Nucl. Phys.}, B, {\bf 650}, 391  \\\ [arXiv: hep-ph/0206071]
\bibitem{heat1} S. Coutu {\it et al.}, 1999, {\it Astropart. Phys.}, {\bf 11}, 429 \\\ [arXiv: astro-ph/9902162]
\bibitem{heat2} J.J. Beatty {\it et al.}, 2004, {\it Phys. Rev. Lett.}, {\bf 93}, 241102 \\\ [arXiv: astro-ph/0412230]
\bibitem{ams1}[AMS Collaboration] J. Alcaraz {\it et al.}, 2000, {\it Phys. Lett.}, B, {\bf 484}, 10
\bibitem{ams2}[AMS Collaboration] M. Aguilar {\it et al.}, 2007, {\it Phys. Lett.}, B, {\bf 646}, 145 \\\ [arXiv: astro-ph/0605254]

\bibitem{Maurin:2002}
  F.~Donato, N.~Fornengo, D.~Maurin, P.~Salati and R.~Taillet,
 D.~Maurin, R.~Taillet, F.~Donato, P.~Salati, A.~Barrau and G.~Boudoul
  [arXiv:astro-ph/0212111]

\bibitem{bess1}
   S.~Orito {\it et al.}  [BESS Collaboration],
   Phys.\ Rev.\ Lett.\  {\bf 84}, 1078 (2000)
   [arXiv:astro-ph/9906426].

\bibitem{bess2}
   T.~Maeno {\it et al.}  [BESS Collaboration],
positive
   Astropart.\ Phys.\  {\bf 16}, 121 (2001)
   [arXiv:astro-ph/0010381].

\bibitem{caprice}
   M.~Boezio {\it et al.}  [WiZard/CAPRICE Collaboration],
   Astrophys.\ J.\  {\bf 561}, 787 (2001)
   [arXiv:astro-ph/0103513].

{
\bibitem{ams98}
   M.~Aguilar {\it et al.} [AMS Collaboration],
Space Station. I:
   Phys.\ Rep.\ {\bf 366}, 331 (2002),
   Erratum-ibid.\ {\bf 380}, 97 (2003).
}

\end{thebibliography}
\end{document}